\documentclass[twocolumn, tradiabstract]{aa} % for a paper on 1 column
\def\degree{\hbox{$^\circ$}}

\def\kms{\hbox{km$\;$s$^{-1}$}}

\usepackage{natbib}
\usepackage{graphicx}
\usepackage{color}
\usepackage[normalem]{ulem}
\begin{document}

\title{Rayleigh--Taylor instability in prominences from numerical simulations including partial ionization effects}

\author{E. Khomenko\inst{1,2,3}, A. D\'{\i}az\inst{1,2}, A. de Vicente\inst{1,2}, M. Collados\inst{1,2}, M. Luna\inst{1,2} }

\institute{Instituto de Astrof\'{\i}sica de Canarias, 38205 La Laguna, Tenerife, Spain \and 
Departamento de Astrof\'{\i}sica, Universidad de La Laguna, 38205, La Laguna, Tenerife, Spain \and 
Main Astronomical Observatory, NAS, 03680, Kyiv, Ukraine\\ \email{khomenko@iac.es} \\
}

\date{Received 25th of October, 2013; accepted 18th of March, 2014}

\abstract{We study the Rayleigh--Taylor instability (RTI) at a prominence--corona transition region in a non-linear regime. Our aim is to understand how the presence of neutral atoms in the prominence plasma influences the instability growth rate, and the evolution of velocity, magnetic field vector and thermodynamic parameters of turbulent drops. We perform 2.5D numerical simulations of the instability initiated by a multi-mode perturbation at the corona--prominence interface using a single-fluid MHD approach including a generalized Ohm's law. The initial equilibrium configuration is purely hydrostatic and contains a homogeneous horizontal magnetic field forming an angle with the direction in which the plasma is perturbed. We analyze simulations with two different orientations of the magnetic field. For each field orientation we compare two simulations, one for the pure MHD case, and one including the ambipolar diffusion in the Ohm's law (AD case). Other than that, both simulations for each field orientation are identical. The numerical results in the initial stage of the instability are compared with the analytical linear calculations. We find that the configuration is always unstable in the AD case. The growth rate of the small-scale modes in the non-linear regime is up to 50\% larger in the AD case than in the purely MHD case and the average velocities of flows are a few percent larger. Significant drift momenta are found at the interface between the coronal and the prominence material at all stages of the instability, produced by the faster downward motion of the neutral component with respect to the ionized component. The differences in temperature of the bubbles between the ideal and non-ideal case are also significant, reaching 30\%. There is an asymmetry between large rising bubbles and small-scale down flowing fingers, favoring the detection of upward velocities in observations.}


\authorrunning{Khomenko et al.}
\titlerunning{RTI in prominences}

\maketitle

%________________________________________________________________
\section{Introduction}

Solar prominences are composed of cool, dense, and partially ionized plasma and their large-scale  magnetic structure remains stable for days, or even weeks, in the solar corona. Prominence material is believed to be supported by the magnetic field  \citep[see reviews by][]{Tandberg-Hanssen1995, Mackay+etal2010}. There are several large-scale models that address the problem of the global stability of prominences, and of the origin of their mass  that may explain observational properties \citep{Pneuman1983, vanBallegooijen1989, Priest+etal1989, Antiochos+Klimchuk1991, Rust+Kumar1994, Antiochos1994, Aulanier1998, DeVore2000, Gibson2006, Aulanier+etal2006}.

On the top of the global stability, prominences are extremely dynamical at small scales, showing a variety of shapes, moving with vertical and horizontal threads \citep{Berger2010, Ryutova+etal2010}. The dynamical appearance of prominences depends on whether these are located above active or quiet regions, and on the relative orientation with respect to the observer. Quiescent prominences usually occur in quiet regions at high latitudes. They usually reach larger heights than active region prominences, and often show very characteristic vertical threads of less than 1 Mm size, and downflow velocities of about 10--20 \kms\ along them\footnote{Due to their peculiar appearance, quiescent prominences with vertical threads are also known as `hedgerow prominences''.}. The origin of these downflowing drops has been addressed in many studies \citep{Gilbert2002, Gilbert2007, Ballegooijen2010, Haerendel2011}, aiming to explain the apparent material motion across the magnetic field lines, the speed of the drops, and the amount of the mass loss and gain.

\citet{Berger2010} find large-scale 20--50 Mm arches, expanding from the underlying corona into the prominences. At the top of these arches, at the prominence--corona transition region (PCTR), there are observed dark turbulent upflowing channels of 4-6 Mm maximum width with a profile typical of the Rayleigh--Taylor (RT) and Kelvin--Helmholtz (KH) instabilities. The upflows rise up to 15--50 Mm, with an average speed of 13--17 \kms, decreasing at the end. Lifetimes of the plumes are about 300--1000 sec. These numbers fit well into the theoretical predictions from the classical theory of plasma instabilities \citep{Isobe+etal2005, Berger+etal2008, Heinzel+etal2008, Ryutova+etal2010, Berger2010, Berger+etal2011} and can therefore be used to derive plasma parameters of the prominences \citep{Hillier+etal2012c}. Plumes and spikes are seen at any time and any possible orientation of the limb portion of a prominence, but they are specially evident in quiescent hedgerow prominences. An alternative explanation for the upflowing plumes was recently proposed by \citet{Dudik+etal2012}, as due to the presence of separatrix layers and reconnection, arguing that RTI can not happen for the magnetic field orientation and plasma parameters expected for prominences. 

From the theoretical point of view, the existence of the instabilities at the PCTR is easily explained since the two media have clearly different densities, temperatures and relative velocities. The analytical linear MHD theory of these instabilities is well developed \citep[see, for example ][]{Chandrasekhar1981, Priest1982}. Numerical simulations in the non-linear regime have been performed for different astrophysical contexts in two and three dimensions \citep{Jun+etal1995a, Jun+etal1995b, Arber2007, Stone+Gardiner2007a, Stone+Gardiner2007b, Isobe+etal2006}. Recent numerical MHD simulations of the RTI in prominences by \citet{Hillier+etal2012a, Hillier+etal2012b}, including a rising buoyant tube in a Kippenhahn--Schl\"{u}ter prominence model show a good agreement with observations.

Prominences are relatively cool and dense objects, with values of temperature and density in the chromospheric range \citep{Tandberg-Hanssen1995}. Therefore, a prominence material is expected to be only partially ionized. The presence of a large number of neutrals must affect the overall dynamics of the plasma, since neutrals do not feel the influence of the magnetic Lorentz force directly, but only through the collisional coupling to ions. The aim of our work is to model the dynamics of the Rayleigh--Taylor instability in the partially ionized prominence plasma in the non-linear regime. The linear theory of the Rayleigh--Taylor and Kelvin--Helmholtz instabilities in a partially ionized plasma has been recently developed by \citet{Soler+etal2012, Diaz+etal2012, Diaz+etal2013}. Different approaches have been used, including a two-fluid and a single fluid modeling. It is known from the ideal MHD that the magnetic field parallel to the perturbation interface stabilizes the system, up to some threshold wavelength $\lambda_c$:
\begin{equation} \label{eq:lambdac}
\lambda_c=\frac{4\pi B_0^2\cos^2\theta}{(\rho_2-\rho_1)g\mu_0}
\end{equation}
where $B_0\cos\theta$ is the value of the magnetic field in the plane of the perturbation, and $\rho_2-\rho_1$ is the density contrast between the two media. For a given value of the magnetic field, perturbations with a wavelength shorter than $\lambda_c$ are stable, while large wavelength perturbations remain to be unstable. The results of the linear theory show that, in a partially ionized plasma, there is no critical wavelength, and perturbations in the whole wavelength range are always unstable \citep{Soler+etal2012, Diaz+etal2012, Diaz+etal2013}. The growth rate of a given perturbation in a partially ionized plasma depends (among other parameters) on the ionization fraction and keeps being rather small for the ionization fractions expected for the prominence plasma.

The linear instability theory has the advantage of being analytical, but then a number of simplifications is necessary, limiting the range of its validity. In order to consider the fully non-linear evolution of the Rayleigh--Taylor instability, and to include more complex physics, numerical simulations have to be performed. Here we report on 2.5D numerical simulations of the RTI at the boundary between the hot corona and a cool prominence taking into account plasma partial ionization by means of the generalized Ohm's law.

%%%%%%%%%%%%%%%%%%%%%%%%%%%%%%%%%%%%%%%%%%%%%%%%%%%%%%%%%%%%%%%%
\begin{figure}[t]
\center
\includegraphics[width=9cm]{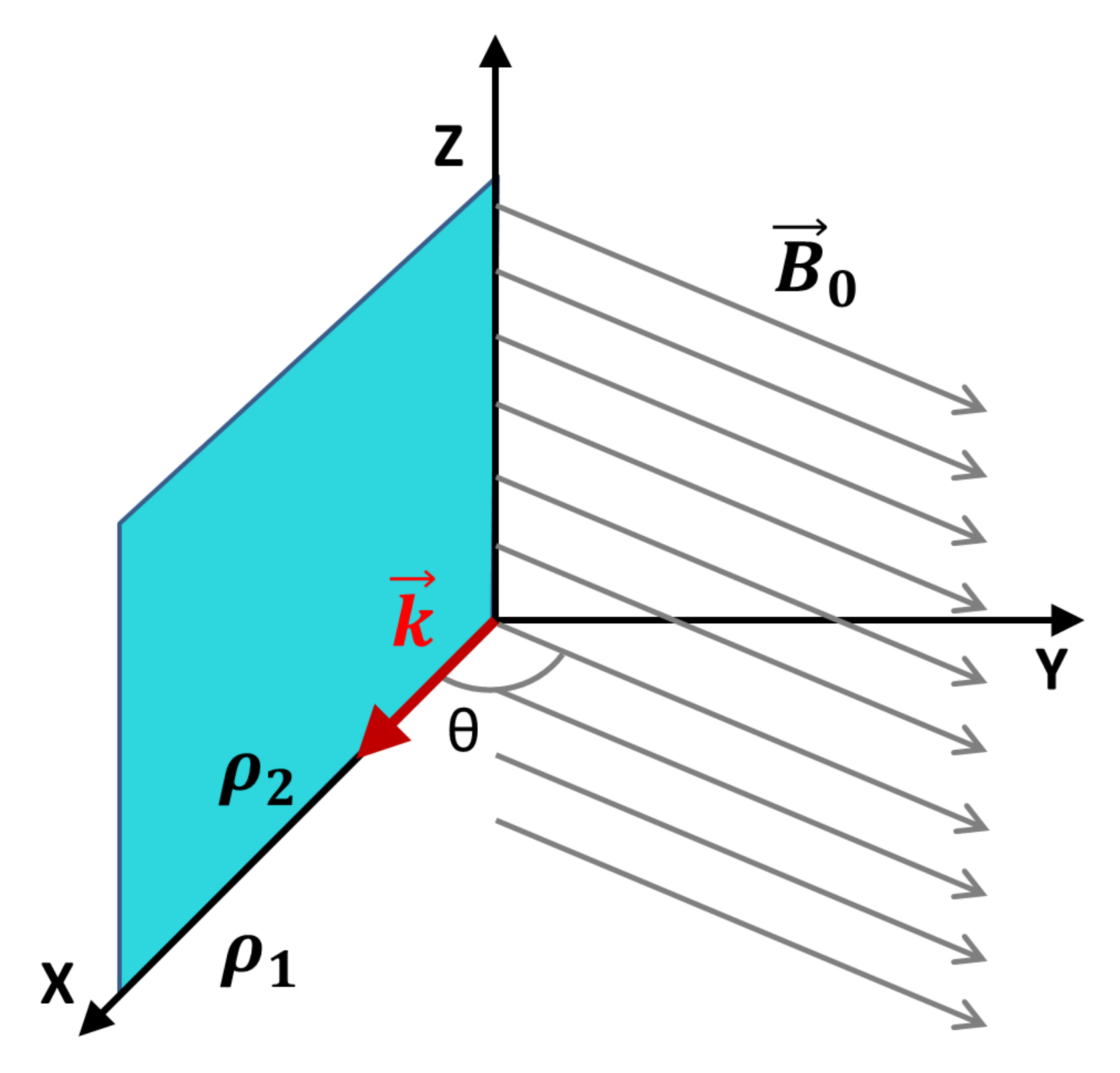}
\caption{{\footnotesize Sketch of the initial configuration. The instability develops in the $XZ$ plane containing the perturbation vector $\vec{k}$; the initial magnetic field $\vec{B}_0$ forms an angle $\theta$ with the $XZ$ plane. $\vec{B}$ is initially lying in $XY$ plane, parallel to the interface separating the prominence and coronal plasma, with $B_z=0$, $B_x=B_0\cos(\theta)$ and $B_y=B_0\sin(\theta)$.} }\label{fig:sketch}
\end{figure}
%%%%%%%%%%%%%%%%%%%%%%%%%%%%%%%%%%%%%%%%%%%%%%%%%%%%%%%%%%%%%%%%

%%%%%%%%%%%%%%%%%%%%%%%%%%%%%%%%%%%%%%%%%%%%%%%%%%%%%%%%%%%%%%%%%%%%%%%%%
\begin{table}[!b]
  \begin{center}
  \centering \caption{ Parameters of the equilibrium configuration, showing the values of temperature ($T$), density ($\rho$), Alfv\'en ($v_a$) and sound ($c_s$) speeds, Ohmic diffusion coefficient ($\eta$, Eq. \ref{eq:etac}), ambipolar diffusion coefficient ($\eta_A$, Eq.~\ref{eq:etaa}), and fraction of neutrals ($\xi_n=\rho_n/\rho$).}
  \label{tab:model}
  \begin{tabular}{ccc}
\hline
                                           & Corona                          & Prominence \\
\hline
$T$ [kK]                              & 400                               &   5  \\
$\rho$ [kg m$^{-3}$]           & 3.7$\times$10$^{-12}$ & 2.9$\times$10$^{-10}$ \\
$v_a$ [km s$^{-1}$]            & 450                               & 53 \\
$c_s$ [km s$^{-1}$]            & 75                                 & 8.3 \\
$\eta$ [m$^2$ s$^{-1}$]      & 7.3                                & 3.3$\times$10$^3$ \\
$\eta_A$ [m$^2$ s$^{-1}$] & 0                                   & 2.3$\times$10$^8$ \\
$\xi_n$                                & 0                                   & 0.9 \\
\hline
 \end{tabular}
  \end{center}
 \end{table}
%%%%%%%%%%%%%%%%%%%%%%%%%%%%%%%%%%%%%%%%%%%%%%%%%%%%%%%%%%%%%%%%%%%%%%%%%%%%%%

%%%%%%%%%%%%%%%%%%%%%%%%%%%%%%%%%%%%%%%%%%%%%%%%%%%%%%%%%%%%%%%%
\begin{figure*}[!ht]
\center
\includegraphics[width=16cm]{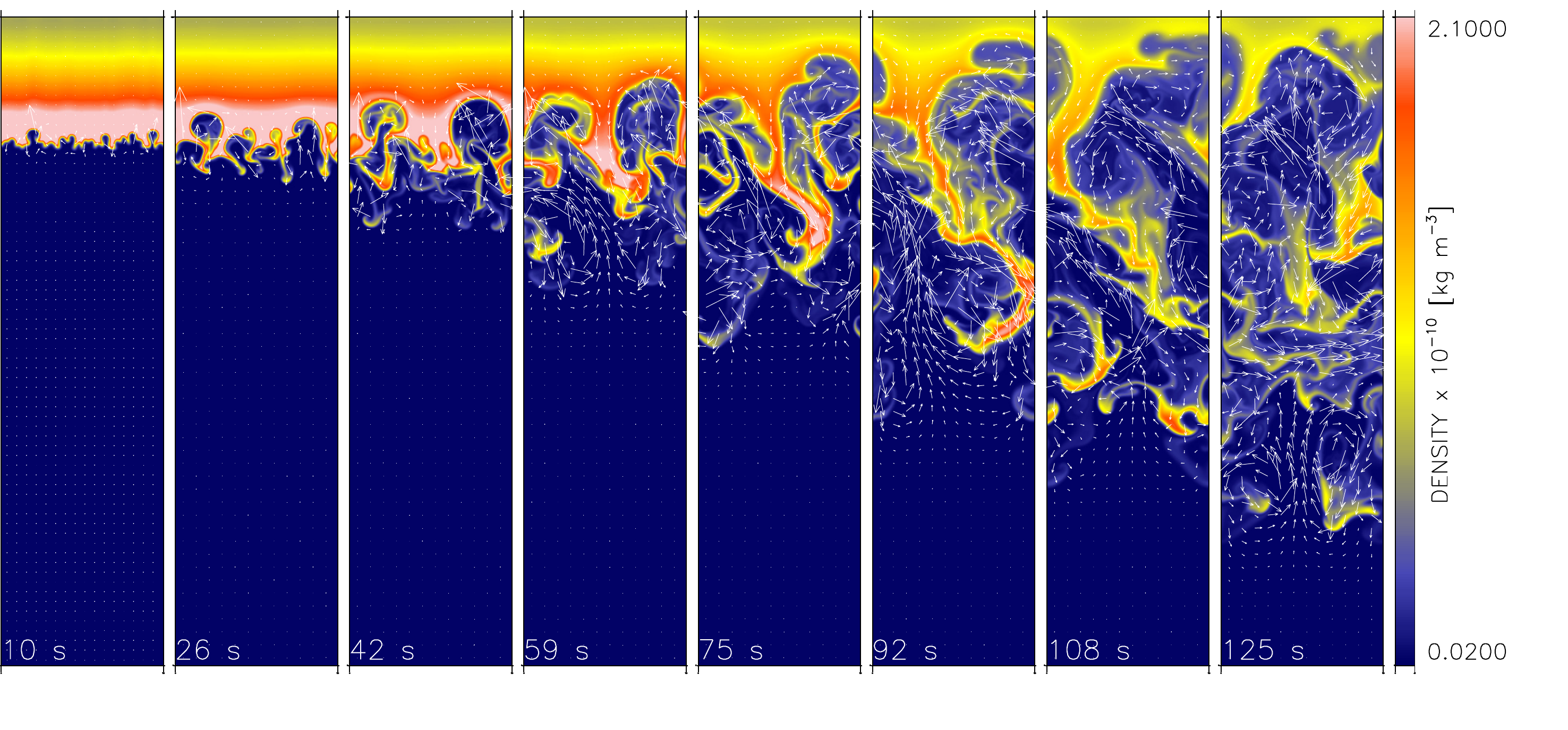}
\includegraphics[width=16cm]{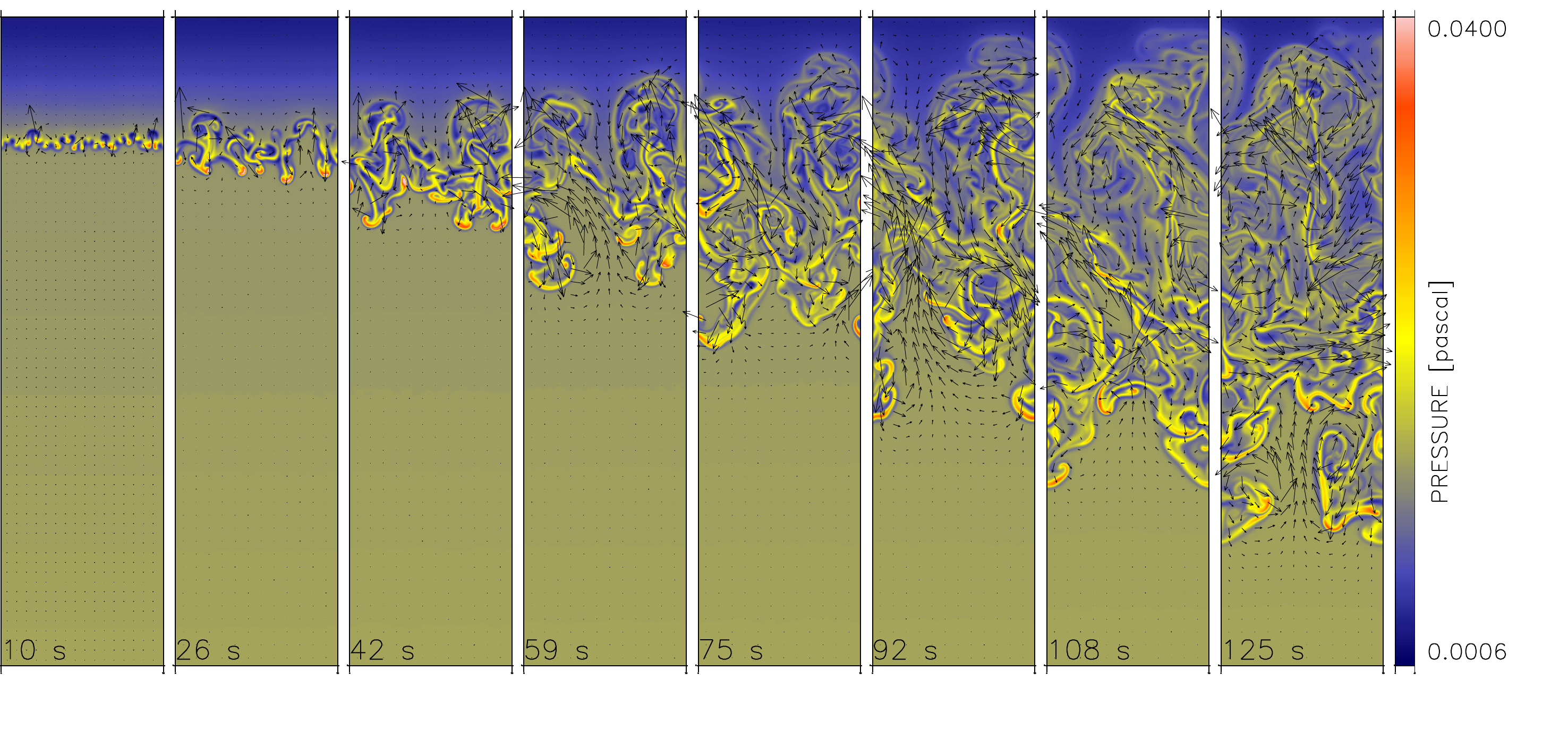}
\caption{{\footnotesize Time evolution of density (top) and pressure (bottom) in the AD simulation with $\theta=90$\degree. The size of each snapshot is 250$\times$1000 km, the elapsed time is given at the bottom of each panel. The velocity field is indicated by arrows. Note the asymmetry between the large-scale rising bubbles and small-scale downflowing fingers in the density images. }
}\label{fig:tevol90}
\end{figure*}
%%%%%%%%%%%%%%%%%%%%%%%%%%%%%%%%%%%%%%%%%%%%%%%%%%%%%%%%%%%%%%%%

%%%%%%%%%%%%%%%%%%%%%%%%%%%%%%%%%%%%%%%%%%%%%%%%%%%%%%%%%%%%%%%%
\begin{figure*}[!ht]
\center
\includegraphics[width=16cm]{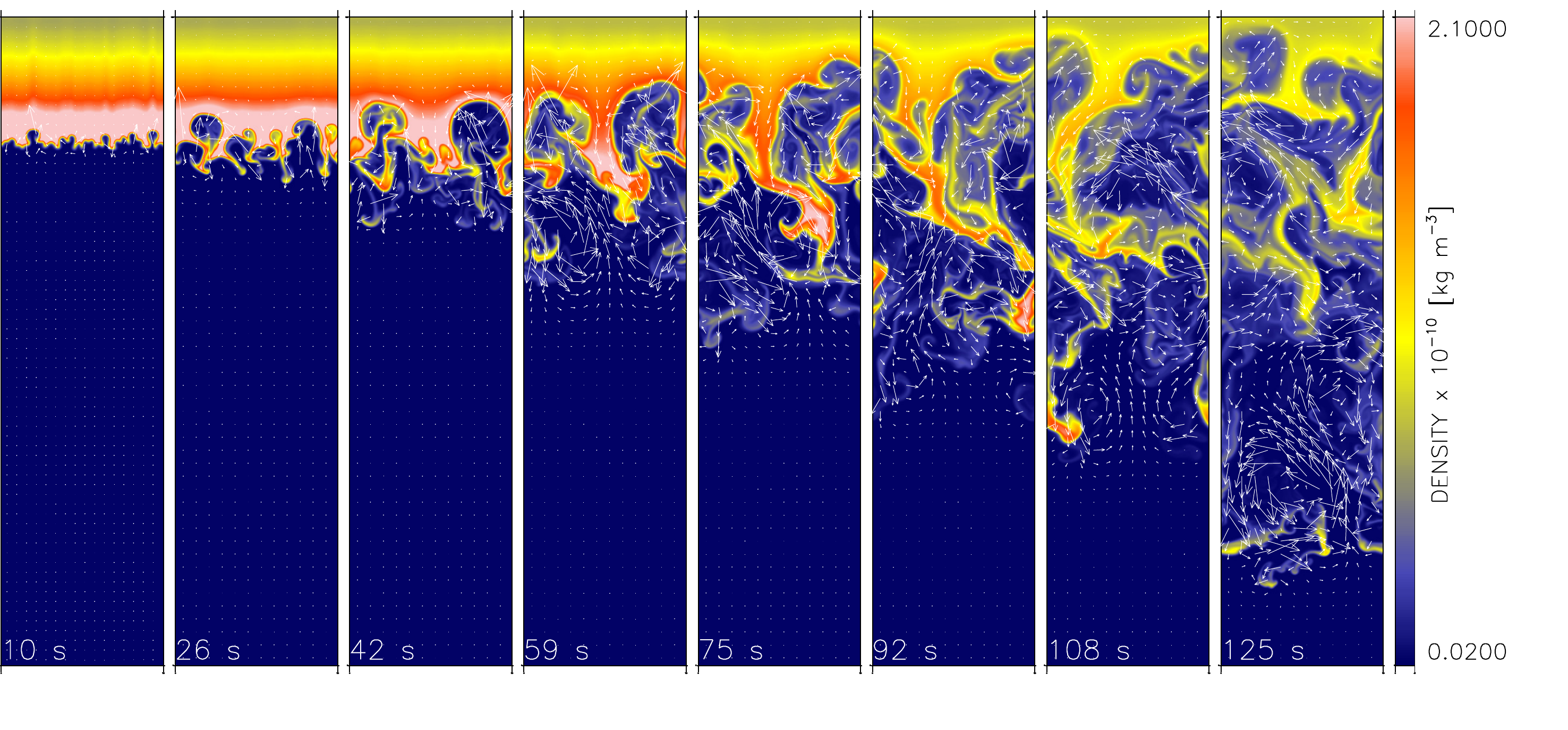}
\includegraphics[width=16cm]{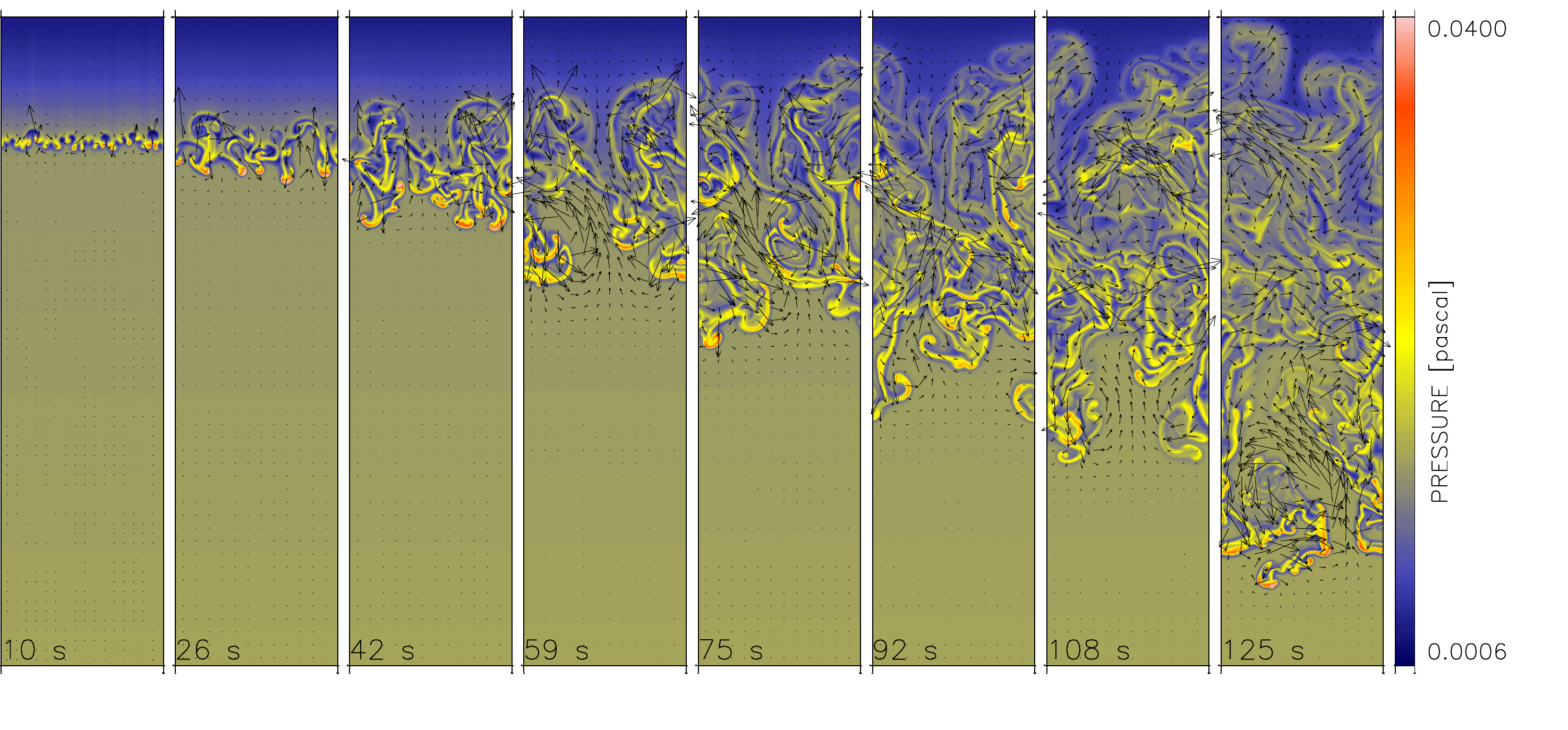}
\caption{{\footnotesize Same as Fig. \ref{fig:tevol90} but for the MHD simulation.}
}\label{fig:tevol90mhd}
\end{figure*}
%%%%%%%%%%%%%%%%%%%%%%%%%%%%%%%%%%%%%%%%%%%%%%%%%%%%%%%%%%%%%%%%

\section{Initial configuration}

Observations of quiescent prominences reveal low field strengths, in the range of 3--30 G \citep{Leroy1989}. The field direction is deduced to be mainly horizontal (making an acute angle with the axis of the prominence) from many observations, though no unique opinion exists in this respect \citep{Mackay+etal2010}.

In the following, we assume the simplest plasma and magnetic configuration, summarized in Fig.~\ref{fig:sketch}. We choose initially homogeneous horizontal magnetic field with a strength of $B_0=10$ G, in prominence and corona. We consider two magnetic field orientations, $\vec{B}_0$ at $\theta=90$\degree\ to $X$ axis (i.e. normal to the $XZ$ plane), and $\vec{B}_0$ at $\theta=89$\degree\ to $X$ axis (slightly skewed from the normal to $XZ$ plane). The plasma is most unstable for these field orientations. For the perturbations forming a larger angle $\theta$ with the magnetic field the instability would take longer time to develop, otherwise leading to similar simulation results \citep[see][]{Khomenko+etal2013}.

Since our aim is to study the effects of non-ideal plasma terms due to neutrals, we choose rather high spatial resolution of 1 km in both $X$ and $Z$ directions.  We simulate only a small portion of the interface between the prominence and the corona, with a size of 250$\times$1000 km. The equilibrium magnetic field is homogeneous and does not influence the force balance. Pressures and densities are obtained from the hydrostatic equilibrium condition, given the temperature stratification. The temperatures of the prominence and corona are initially constant with a smooth transition. This equilibrium is different from what is usually assumed for prominences, where the magnetic field is expected to exert a force on the plasma and prevents it from falling. Despite this, since we only simulate a small portion of the interface, we expect that the effects of the curvature of the equilibrium magnetic field are not significant. The results presented here represent an initial exploratory study and such a configuration allows us to make a direct comparison with the linear theory where a similar configuration is also assumed \citep{Diaz+etal2013}.

Despite the advantage of its simplicity, the homogeneous current-free field configuration is unfavorable for the manifestation of non-ideal plasma effects. The importance of these effects depends on currents (terms proportional to  $\eta_A\vec{J}_\perp$ in the Eqs. \ref{eq:system} below). As the instability is being developed, the currents ($\vec{J}_\perp$) appear only due to perturbations in the magnetic field, at the interface between the two media, and their influence is limited to a small region (PCTR), unlike in a non-potential magnetic field configuration.  This lack of currents is one of the reasons why partial ionization effects are relatively local and were not considered so far in theoretical studies. On the other hand, the ionization degree of the prominence plasma can be very low because of its low temperature, and the large fraction of neutrals may partially compensate for the weakness of  currents and increase to significant values the diffusion coefficient due to neutrals, i.e. $\eta_A$, so the terms proportional to $\eta_A\vec{J}_\perp$ become significant. Table~\ref{tab:model} summarizes the parameters of our equilibrium configuration. For these parameters, Eq.~\ref{eq:lambdac} gives the critical wavelength $\lambda_c=38$ km for $\theta$=89\degree. There is no critical wavelength in the $\theta=90$\degree\ case.

%%%%%%%%%%%%%%%%%%%%%%%%%%%%%%%%%%%%%%%%%%%%%%%%%%%%%%%%%%%%%%%%
\begin{figure}
\center
\includegraphics[width=7cm]{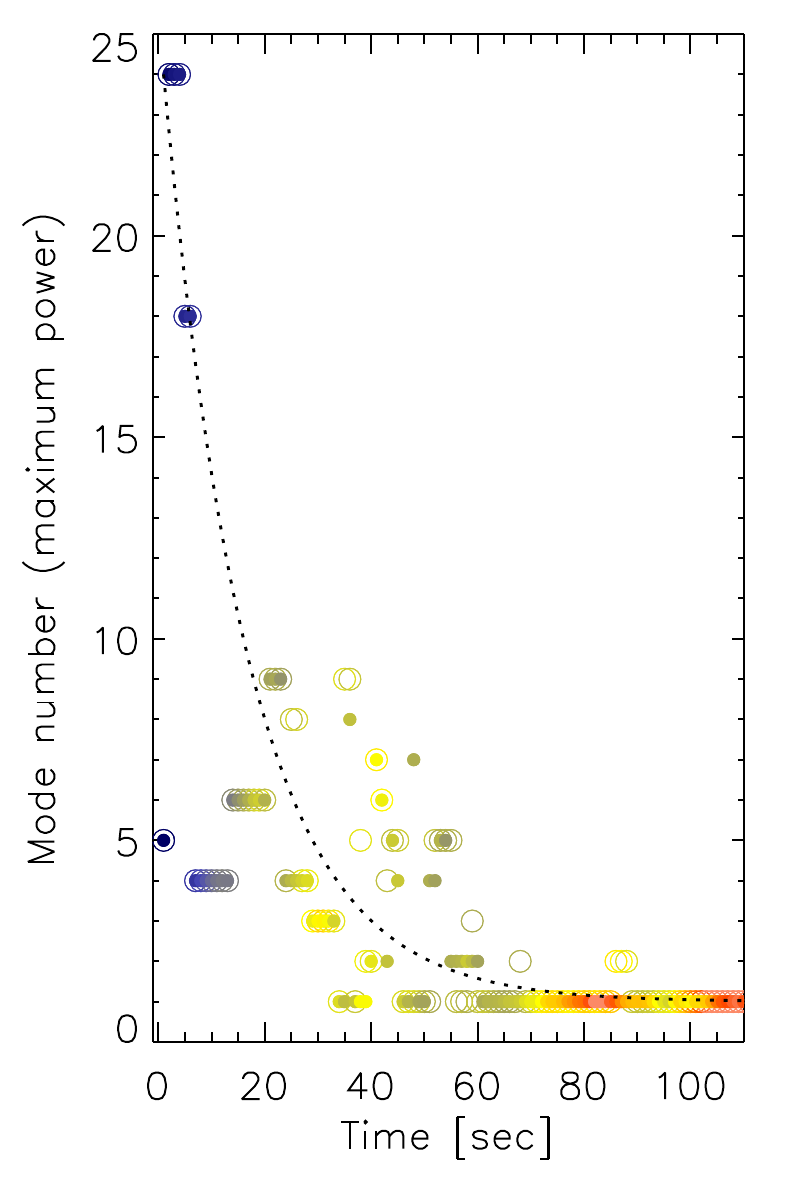}
\caption{{\footnotesize Mode number (m, see Eq. \ref{eq:mmode})  with maximum Fourier amplitude of the pressure perturbation in the  AD (filled circles) and MHD (open circles) simulations with $\theta=90$\degree. The wavelength of the modes varies from 10 km (mode 25) to 250 km (mode 1). The amplitude of the modes is indicated by color from blue (smaller) to red (larger). Solid line is an exponential function $\sim\exp(-t/16)$ shown for illustration purposes. }}\label{fig:maxscale}
\end{figure}
%%%%%%%%%%%%%%%%%%%%%%%%%%%%%%%%%%%%%%%%%%%%%%%%%%%%%%%%%%%%%%%%

\section{Numerical solution}

We solve numerically the non-ideal-MHD equations of conservation of mass, momentum, internal energy, and the induction equation \citep{Khomenko+Collados2012}:

\begin{eqnarray} \label{eq:system}
\frac{\partial \rho}{\partial t} & =& - \vec{\nabla}\left(\rho\vec{u}\right)  \\ \nonumber
\rho\frac{D\vec{u}}{D t}&  =& \vec{J}\times\vec{B} + \rho\vec{g} - \vec{\nabla}p \\ \nonumber
\frac{D p}{D t} & =&  -\gamma p\vec{\nabla}\vec{u} + (\gamma-1)(\eta\mu_0\vec{J^2} + \eta_A\mu_0 \vec{J_{\bot}^2})\\ \nonumber
\frac{\partial\vec{B}}{\partial t} & = &\vec{\nabla}\times \left[ (\vec{u}\times\vec{B}) - \eta\mu_0\vec{J} - \eta_A\mu_0\vec{J_{\bot}} \right]
\end{eqnarray}
where the gravity is equal to $g=2.74\times 10^2$ m s$^{-2}$, the adiabatic index $\gamma=5/3$. The variables $\rho$, $p$ are summed over the plasma components (electrons ($e$), ions ($i$) and neutrals ($n$)), the center of mass velocity $\vec{u}$ is an average over the velocity of individual species:
\begin{equation} \label{eq:u} 
\vec{u}=\frac{\sum_{\alpha=e,i,n}\rho_\alpha \vec{u}_\alpha }{\rho}\approx\frac{\rho_i\vec{u}_i + \rho_n\vec{u}_n}{\rho} \end{equation}
and $\vec{J}_{\perp}$ is the component of the current perpendicular to the magnetic field:
\begin{equation}
\vec{J}_{\perp} = -\frac{[\vec{J}\times\vec{B}]\times\vec{B}}{|B|^2}
\end{equation}

In these equations we neglect the non-diagonal components of the pressure tensor and assume small drift velocities $\vec{w}_{\alpha}=\vec{u} - \vec{u}_{\alpha}$ ($\alpha=e,i,n$),  i.e. $|w_\alpha | \ll |u|, |u_\alpha |$ (see Section \ref{sect:drift} for the discussion of the value of the ion-neutral drift momentum in our simulations). The latter condition allows to remove the terms containing $w_{\alpha}^2$.  In the Ohm's law we neglected the time variation of the relative ion-neutral drift velocity $\vec{u}_i - \vec{u}_n$. This can be safely done since the expected instability growth rate is significantly below the ion-neutral collisional frequency $\nu_{in} \sim 10^4$ s$^{-1}$, \citep{Diaz+etal2013}. We only keep Ohmic and ambipolar\footnote{see Appendix A for the discussion of the common nomenclature for this term used in the literature. } diffusion terms. The linear analysis reveals that for the typical conditions of prominence plasma the ambipolar diffusion term dominates by orders of magnitude over the neglected Hall and battery terms \citep{Diaz+etal2013}.  The ambipolar diffusion coefficient is equal to \citep{Braginskii1965}:
\begin{equation}
\label{eq:etaa}
\eta_A = \frac{(\rho_n/\rho)^2|B|^2}{(\rho_i\nu_{in} + \rho_e\nu_{en})\mu_0}
\end{equation}
The Ohmic diffusion coefficient is given by:
\begin{equation} \label{eq:etac}
\eta=\frac{m_e(\nu_{ei} + \nu_{en})}{e^2n_e\mu_0}
\end{equation}
and theoretical values of the collisional frequencies are \citep{Spitzer1968, Braginskii1965}:
\begin{eqnarray} \label{eq:nus}
\nu_{in}&=&n_{n}\sqrt{\frac{8 k_B T}{\pi m_{in}}}\sigma_{in} \\ \nonumber
\nu_{en}&=&n_{n}\sqrt{\frac{8 k_B T}{\pi m_{en}}}\sigma_{en} \\ \nonumber
\nu_{ei} &=& \frac{e^4 n_e \Lambda}{3\epsilon_0^2 m_e^2}\left(\frac{m_e}{2\pi k_B T}\right)^{3/2} 
\end{eqnarray}
where $e$ is the electron charge, $n_e$ is the electron number density, $m_{in}=m_i m_n/(m_i + m_n)$ and $m_{en}=m_e m_n/(m_e + m_n)$. The respective cross sections are $\sigma_{in}=5\times10^{-19}$ m$^2$ and $\sigma_{en}=10^{-19}$ m$^2$. The  Coulomb logarithm $\Lambda$ for $T<50$ eV ($T< 600.000$ K) is equal to:
\begin{equation}
\Lambda=23.4 - 1.15\log_{10}{n_e}+3.45\log_{10}{T}
\end{equation}

Equations \ref{eq:system} were solved by means of our code {\sc mancha} \citep{Khomenko+etal2008, Felipe+etal2010, Khomenko+Collados2012} with the inclusion of the physical ambipolar diffusion term in the equation of energy conservation and in the induction equation. We evolve the electron number density $n_e$ in time by the Saha equation. The simulations are done in 2.5D, meaning that the derivatives are done only in $X$ and $Z$ directions, but the vector quantities (velocity and magnetic field) are allowed to have three components. We use periodic horizontal boundary conditions and closed vertical boundary conditions.

The code  {\sc mancha} evolves (non-linear) perturbations to an equilibrium state. In order to initiate the instability in the current experiment we proceeded in the following way. We constructed two models: (1) model\#1 represents a horizontally homogeneous coronal model in hydrostatic equilibrium at coronal temperature $T_{\rm cor}$ from Table \ref{tab:model}; (2) model\#2 is a model of corona with prominence on the top. The vertical hydrostatic equilibrium in model\#2 is imposed in each column with temperature varying from $T_{\rm cor}$  (at the lower part of the domain) to $T_{\rm prom}$ (at the upper part) and the corona-prominence interface corrugated according to Eq~\ref{eq:mmode}. The difference between the model\#2 and model\#1 was used as a perturbation to initiate the instability. The position $z$ where the temperature changes from coronal to prominence values in the model\#2 varies with horizontal location $x$, otherwise the perturbation is in equilibrium and does not evolve. We used the multi-mode perturbation to the corona-prominence interface from \citep{Jun+etal1995b}:
\begin{eqnarray} \label{eq:mmode}
\delta z(x) = z(x) -  z_i &= & \sum_{m=-M}^{m=M} a_m \cos(\frac{2\pi m x}{L} + \varphi_m) \\ \nonumber
& + & b_m \sin(\frac{2\pi m x}{L} + \varphi_m) 
\end{eqnarray}
where $L=250$ km and $a_m, b_m,\varphi_m$ are randomly generated amplitudes and phases, and $z_i=800$ km is the location of the interface. The amplitudes $a_m$ and $b_m $vary in the range 0--10 km. We used $M=25$ modes from 10 to 250 km in wavelength. In the $\theta$=89\degree\ simulation, under ideal MHD conditions, three quarters of these modes with $\lambda<\lambda_c$ are expected to be damped due to magnetic field stabilization effects.

Our code uses hyperdiffusivity for stabilizing the numerical solution.  One has to be extremely careful with the choice of the hyperdiffusivity amplitude and model since hyperdiffusive terms are numerical analogs of the physical dissipative terms, such as viscosity, conductivity and Ohmic diffusion \citep{Stein+Nordlund1998,Caunt+Korpi2001,Vogler2005, Felipe+etal2010} and their action may resemble the action of the physical terms. In particular, the numerical diffusive terms are able to amplify the growth rate of small-scale modes of the RTI, the same as the physical diffusion would do. To avoid this problem in our simulations, we kept the amplitude of artificial hyperdiffusive terms 2-3 orders of magnitude smaller that the physical ambipolar diffusion. This way we have assured that the characteristic time scales of action of the hyperdiffusive terms are orders of magnitude larger than that of the physical ambipolar diffusion term and their influence is not significant during the temporal interval covered by the simulations.

In the following we will refer to the simulations with ambipolar diffusion term ``on'' as ``AD'' simulations; and the ones with the ambipolar term ``off'' as MHD simulations.

%%%%%%%%%%%%%%%%%%%%%%%%%%%%%%%%%%%%%%%%%%%%%%%%%%%%%%%%%%%%%%%%
\begin{figure}
\center
\includegraphics[width=9cm]{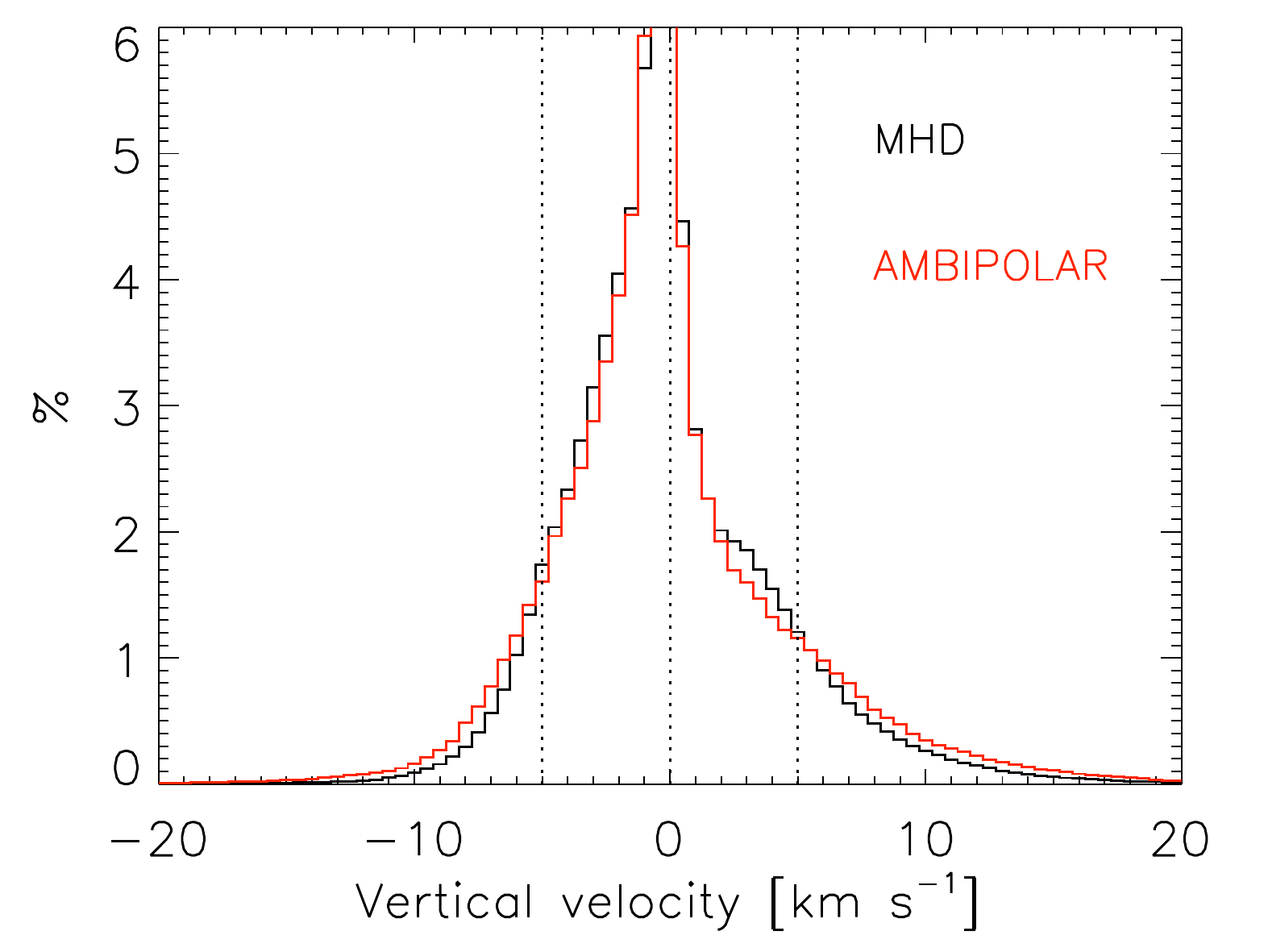}
\caption{{\footnotesize Velocity histograms over the whole simulation time for the $\theta=90$\degree\ MHD and AD runs. Note the asymmetry of the histogram and the slight tendency of the AD simulation to develop larger velocities.} }\label{fig:vz90}
\end{figure}
%%%%%%%%%%%%%%%%%%%%%%%%%%%%%%%%%%%%%%%%%%%%%%%%%%%%%%%%%%%%%%%%

%%%%%%%%%%%%%%%%%%%%%%%%%%%%%%%%%%%%%%%%%%%%%%%%%%%%%%%%%%%%%%%%
\begin{figure}
\center
\includegraphics[width=9cm]{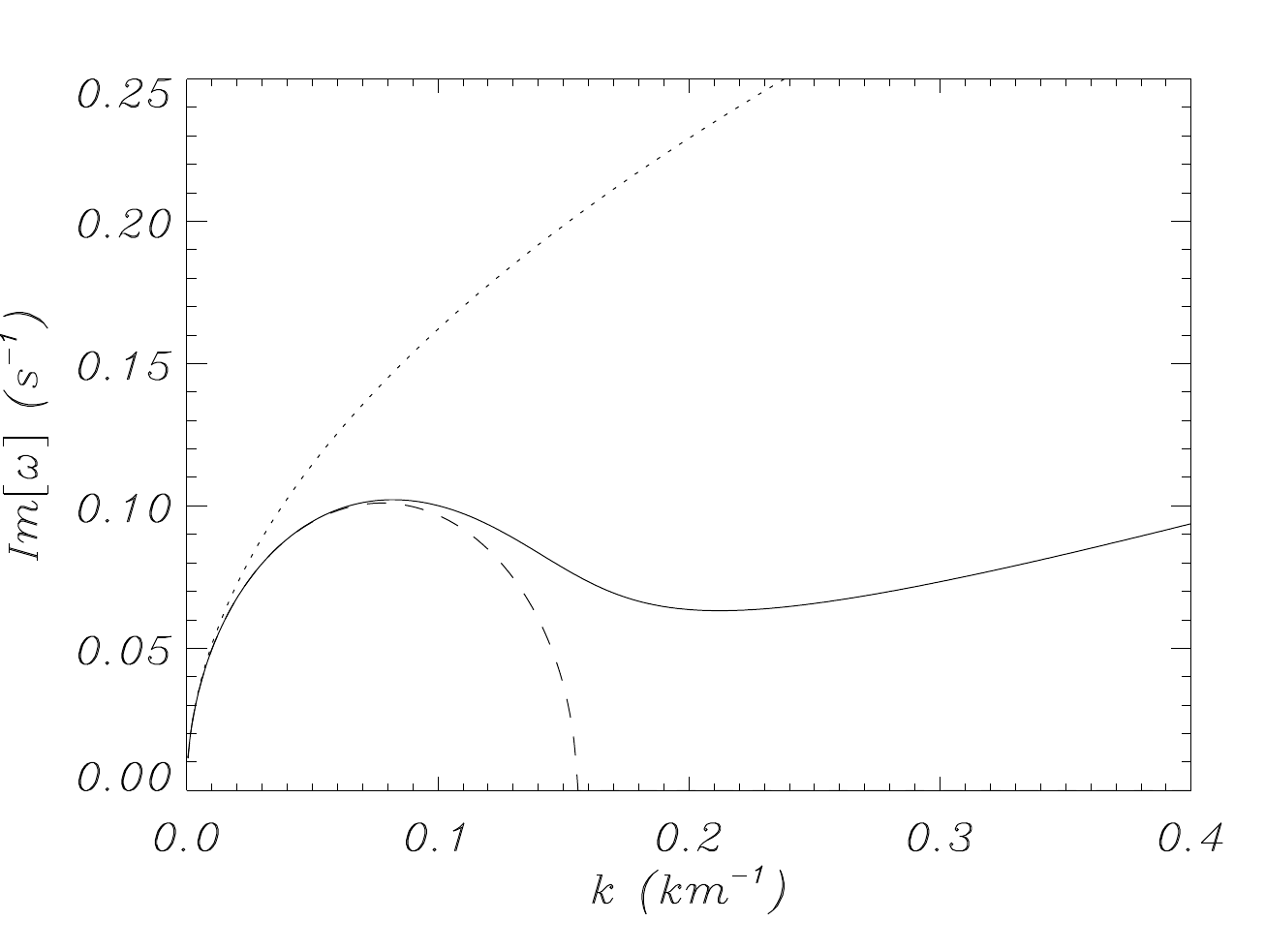}
\caption{{\footnotesize Growth rate of the RTI from the linear theory according to \cite{Diaz+etal2013}, as a function of the wave number of the perturbation along the discontinuity. The parameters of the atmosphere assumed for the linear calculation are those from Tab.~\ref{tab:model}, the magnetic field is inclined with respect to $\vec{k}$ by $\theta=89$\degree\ (dashed line) and by $\theta=90$\degree\ (dotted line). Solid line shows the case of $\theta=89$\degree\ with ambipolar diffusion term taken into account. }
}\label{fig:linear}
\end{figure}
%%%%%%%%%%%%%%%%%%%%%%%%%%%%%%%%%%%%%%%%%%%%%%%%%%%%%%%%%%%%%%%%

%%%%%%%%%%%%%%%%%%%%%%%%%%%%%%%%%%%%%%%%%%%%%%%%%%%%%%%%%%%%%%%%
\begin{figure*}
\center
\includegraphics[width=14cm]{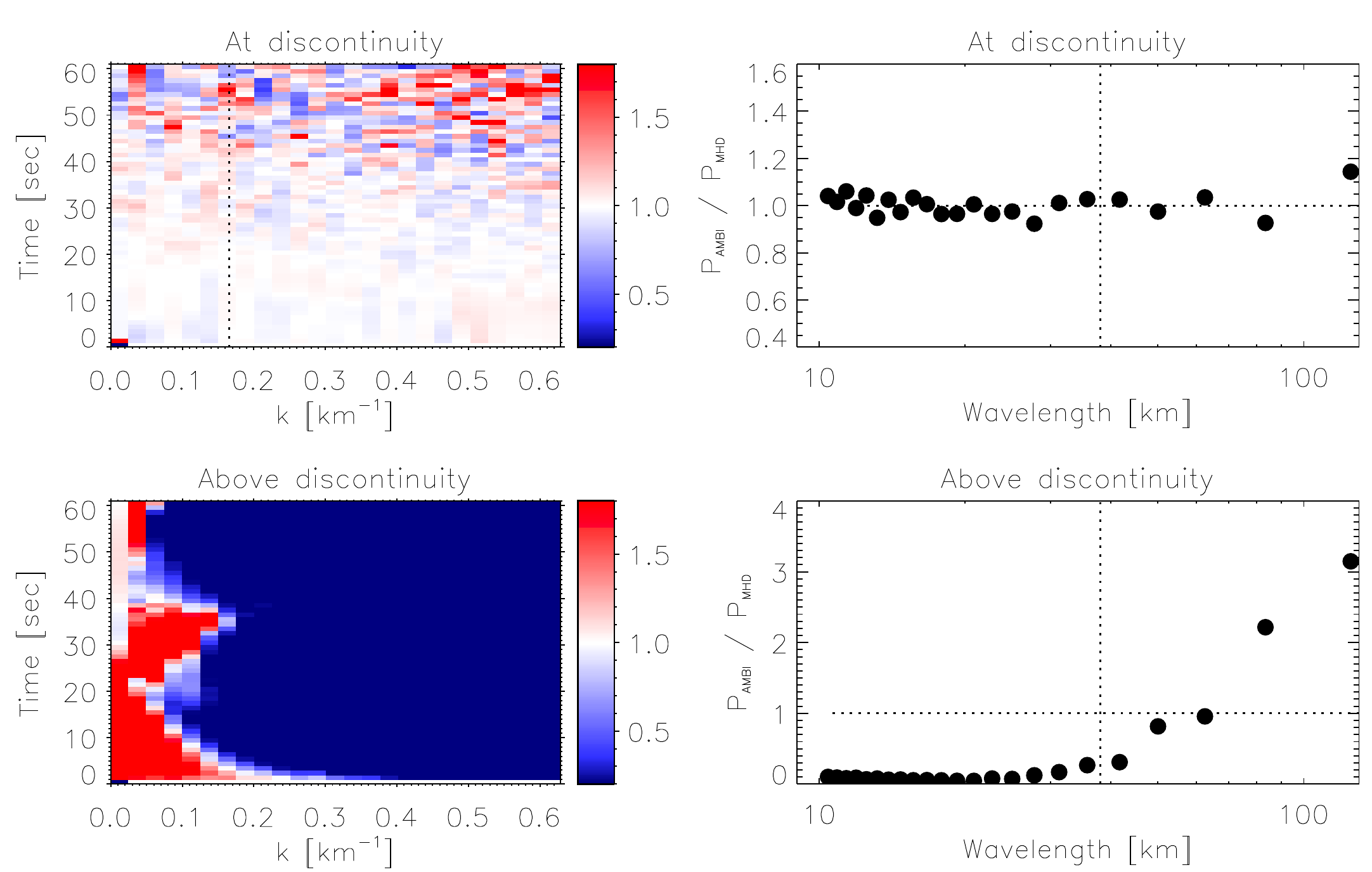}
\caption{{\footnotesize Fourier analysis of scales developed in the $\theta=90$\degree\ simulations. Panels on the left shows the relative power in the AD simulation divided by the power in MHD simulation as a function of horizontal wave number along the discontinuity, $k$, and time. Red colors mean that the AD simulation has more power. Panels on the right give the time average of the power ratio from the panels on the left.} }\label{fig:scales90}
\end{figure*}
%%%%%%%%%%%%%%%%%%%%%%%%%%%%%%%%%%%%%%%%%%%%%%%%%%%%%%%%%%%%%%%%

\section{Results}

\subsection{Simulations with $\vec{B}$ normal to the perturbation plane.}

Figs.~\ref{fig:tevol90} and \ref{fig:tevol90mhd} give the time evolution of density and pressure in the AD and MHD simulations with $\theta=90$\degree. Prior to discussing this simulation we have verified that the behavior of the instability in the non-linear regime follows the well known pattern from previous simulations by other authors. Fig.~\ref{fig:maxscale} shows the dominant spatial scale in the simulations with $\theta=90$\degree\ (both AD and MHD cases are shown) as derived by Fourier transforming the whole domain and calculating the wavelength (mode) of maximum power at each time moment. In this calculation we performed 2D Fourier transformation and then averaged the power corresponding to the vertical direction. Fig.~\ref{fig:maxscale}  can be directly compared with, e.g., \citet{Jun+etal1995b} and shows that large scales tend to dominate in the non-linear stage of the instability since smaller bubbles merge into the larger structures,  in agreement with the classical behavior seen in earlier studies \citep{Youngs1984, Gardner1988}. This tendency is similar in the AD and MHD cases. The situation with $\vec{B}$ normal to the perturbation plane is equivalent to a purely hydrodynamical case as there is no cut-off wavelength. In the linear regime, the growth rate of small-scale modes is faster than of the large-scale modes, see, e.g. \cite{Chandrasekhar1981, Priest1982}. However, in the non-linear regime large-scale modes have larger terminal velocities and with time overtake the small-scale modes and dominate the evolution of the system \citep{Jun+etal1995b, Wang+Robertson1985, Isobe+etal2006, Stone+Gardiner2007a, Stone+Gardiner2007b, Hillier+etal2012a}.

The time evolution of pressure and density in Figs.~\ref{fig:tevol90} and  \ref{fig:tevol90mhd} shows the development of very small scales. We notice the presence of   inverse mushroom shapes (especially evident in the pressure images from 26 to 59 sec), typical for the secondary Kelvin-Helmholtz instability  \citep{Youngs1984, Gardner1988, Jun+etal1995b}. The comparison of the  AD and the MHD simulation reveals a different particular form of the turbulent flows (but statistically equivalent, as we show below), since the ambipolar diffusion, acting on small scales, forces their different evolution. Particularly interesting is the difference in scales between the density and pressure variations. Pressure (bottom panel of Fig.~\ref{fig:tevol90}) is initially maintained constant across the border between the prominence and the corona, forced by the hydrostatic equilibrium condition. However, the pressure scale height in the prominence (upper) part of the domain is small, causing the visible decrease of the pressure toward the upper boundary (blue color). The turbulent flow of the instability mixes up the fluids and tends to smooth the transition. The density variations (upper panel) have a pronounced asymmetry between the upward rising bubbles and downflowing fingers. The material flows down at small scales. This asymmetry is caused by mass conservation. The hot and less dense coronal material rises up and easily expands forming a large bubble. To move the same amount of material, the dense prominence plasma has to occupy a much narrower volume on its downward motion. The density images can be taken as a proxy for the $H_{\alpha}$ observational images. We can speculate that in this type of imaging observations, the upward rising bubbles should be significantly more visible (as dark structures) due to their larger scales, producing a visible asymmetry between the upward and downward moving material.

%%%%%%%%%%%%%%%%%%%%%%%%%%%%%%%%%%%%%%%%%%%%%%%%%%%%%%%%%%%%%%%%
\begin{figure*}
\center
\includegraphics[width=16cm]{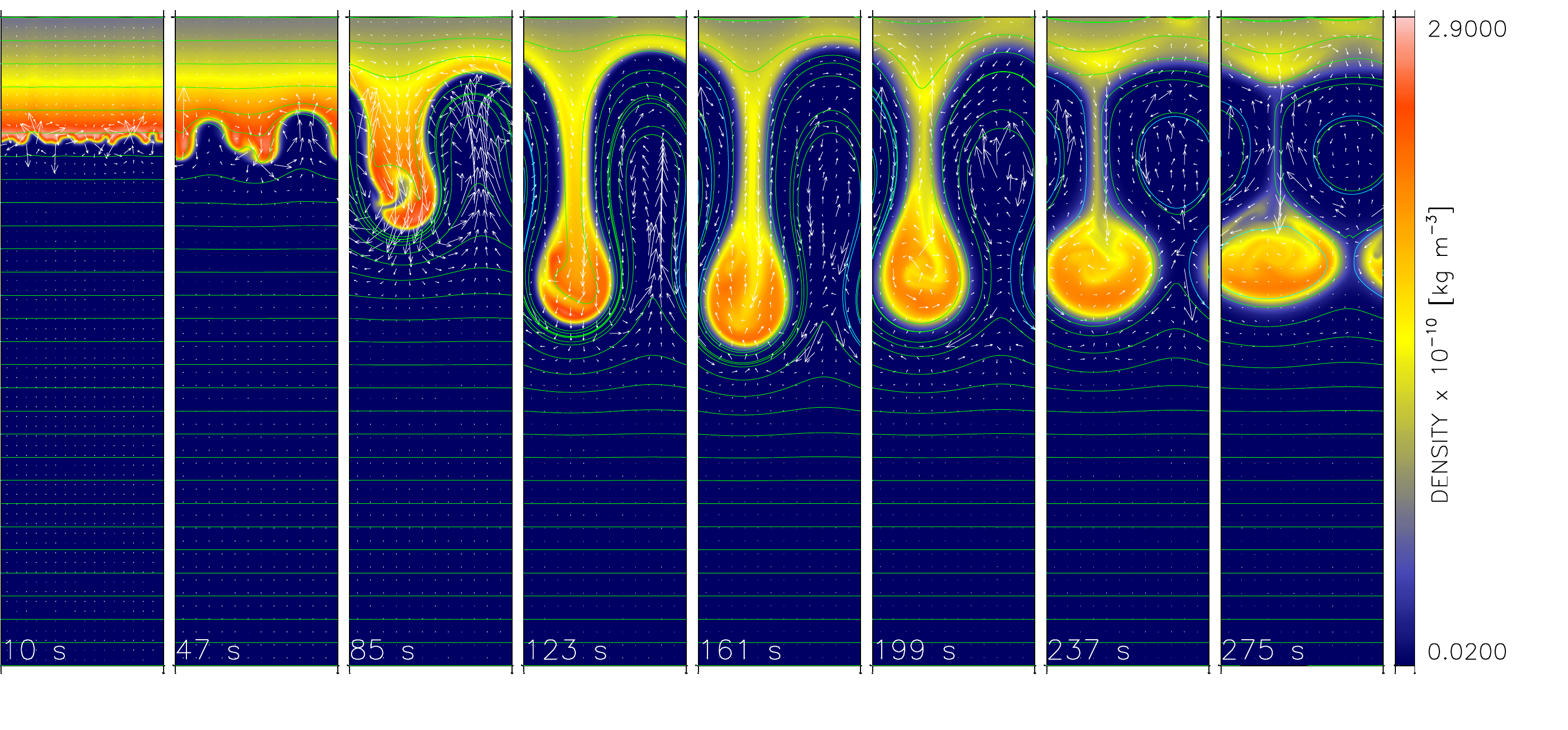}
\includegraphics[width=16cm]{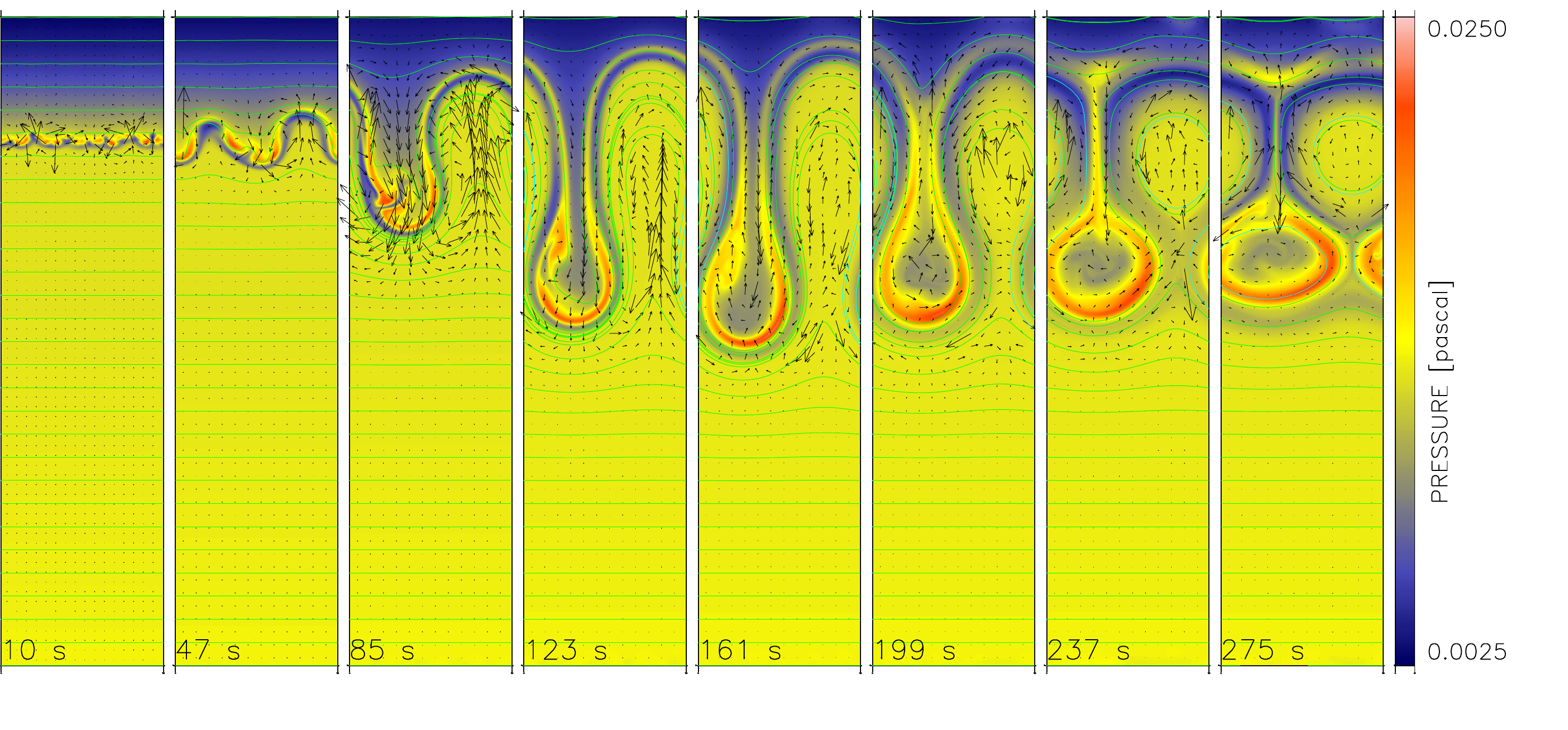}
\caption{{\footnotesize Time evolution of density (top) and pressure (bottom) in the  AD simulation with $\theta=89$\degree. The size of each snapshot is 250$\times$1000 km, the elapsed time is given at the bottom of each panel. Green lines are projections of the magnetic field lines into $XZ$ plane, the velocity field is indicated by arrows. Note the difference in scales compared to the $\theta=90$\degree\ case. } }\label{fig:tevol89}
\end{figure*}
%%%%%%%%%%%%%%%%%%%%%%%%%%%%%%%%%%%%%%%%%%%%%%%%%%%%%%%%%%%%%%%%

Fig.~\ref{fig:vz90} provides a histogram of the vertical velocities (positive means upward motion) around the PCTR\footnote{Here and below we take as PCTR the region from 780 to 940 km, with the largest ambipolar diffusion term.} in the AD simulation with $\theta=90$\degree\ (red line), compared to the MHD simulation (black line).  Both histograms are asymmetric in a way that upflows are fewer (occupying only 1/3 of all points) but stronger, and the average velocity is zero at any time. Note that the prevalence of downflowing points does not contradict the statement that the downflowing regions appear as small-scale fingers on the density images in Figs. \ref{fig:tevol90} and \ref{fig:tevol90mhd},  since the less dense coronal material around the fingers is also involved in the downflowing motion, but is not apparent on the density images. There is a slight, but indicative, difference between the AD and the MHD histograms, particularly apparent in their wings. The red line goes systematically above the black line, meaning that there are more cases of extreme velocities in the AD case. To quantify this difference, we measured the percentage of points with the absolute value of velocity above 5 \kms\ obtaining 12\% for the MHD simulation and 16\% for the AD simulation. This result indicates that the RTI in the non-linear regime develops slightly larger velocities when taking into account the influence of the neutral component, compared to the purely MHD case.

As a next step we analyzed the scales developed in the non-linear regime in the simulation with $\theta=90$\degree. As mentioned above, the case of the magnetic field normal to the plane of the instability is analogous to a purely hydrodynamical case since there is not cut-off wavelength. In the linear regime, the small-wavelength perturbation should grow faster, but in the non-linear regime, the large-scale perturbations become more important due to non-linear interaction of the turbulent flows. Indeed, this behavior is reproduced both in the MHD and AD simulation, as can be seen in Fig.~\ref{fig:maxscale}. In the non-linear regime, even for the $\theta$=90\degree\ case, the plasma motion generates currents in the plane of the perturbation. These currents dissipate through the corresponding term in the energy equation and cause diffusion effects, especially affecting small scales in the simulation. The dissipative action of the ambipolar term can not be taken into account in the linear theory since the dissipative term in the energy equation is non-linear and is removed from the linear analysis \citep{Diaz+etal2013}.

The instability growth rate from the linear theory for $\theta=90$\degree\ is given by dotted line in Fig~\ref{fig:linear}, according to the calculations presented in \cite{Diaz+etal2013}. This figure indicates that, depending on the wave number, the time scale of the development of the instability is of the order of tens of seconds. For wavelengths around $\lambda=2\pi/k=100$ km the time scale is $t=2\pi/{\rm Im}(\omega)\approx 50$ sec, and decreasing to $t\approx 25$ sec for $\lambda\approx 30$ km. Figs.~\ref{fig:tevol90} and \ref{fig:tevol90mhd} show that the instability is already outside of the linear regime at about 10 sec of the simulation. Thus, our numerical results are not directly comparable to the linear theory, but we can take the linear theory as an indication.

Fig.~\ref{fig:scales90} shows the Fourier analysis of scales of the AD simulation, relative to the MHD simulation. In view of the above considerations we have selected two zones of the domain, one of them around the discontinuity (PCTR), and another one above it. The ambipolar diffusion coefficient is only significant in the prominence part of the domain above the discontinuity, and therefore we do not consider the coronal part of the domain for the Fourier analysis. At each time moment we calculated the power as a function of horizontal wave number $k$ along the discontinuity by 2D Fourier-transform in space of the corresponding zone of the snapshot of pressure variations (as the variable that develops more fine structures), and by averaging the power for the vertical wave number to decrease the noise.  At each time moment we re-normalized the total power to maintain it constant in time. This way we obtain a power as a function of horizontal wavelength and time at two domains, at and above the discontinuity.  This procedure was applied separately to the AD and MHD simulation. We then divided the AD power map by the corresponding MHD map in each domain. The result is given at the left panels of Fig.~\ref{fig:scales90}.

Above the discontinuity (bottom left panel in Fig.~\ref{fig:scales90}), the ambipolar diffusion term indeed acts as a pure diffusion quickly removing all small scales and amplifying large scales relative to them. At the PCTR, the behavior of Fourier harmonics is strikingly different. There we do not observe any significant change in power between the AD and MHD cases. The relative power fluctuates in time, but the average remains around one for all harmonics. This result indicates that the ambipolar term can act in a clearly different way in situations where the instability is being developed at PCTR and outside this region.

\subsection{Simulations with $\vec{B}$ at 89\degree\ to the perturbation plane.}

The time evolution of density and  pressure in the AD simulation with $\vec{B}$ skewed from the normal direction is given in Fig.~\ref{fig:tevol89}\footnote{We do not show the corresponding MHD simulation in the same format since the developed structures are similar and the differences are much better appreciated after the Fourier analysis or just by subtracting the individual snapshots, see Figures \ref{fig:vzlinear89}, \ref{fig:scales89} and \ref{fig:dt89}.}. By just rotating the field by 1\degree , the scales developed in the simulation have completely changed: the small scales disappear and only one big drop is developed after about 100 sec of the simulation. The small scales can not develop the instability because of the cut-off induced by the magnetic field, $\lambda_c\approx 38$ km for the parameters of our atmospheric model in equilibrium. While the growth rate of small scale modes ($\lambda < \lambda_c$) is not zero as in the purely MHD case, it  is still smaller compared to large scales, see Figure 5 in \cite{Diaz+etal2013}. One can appreciate from the figure that at $t\sim10$ sec the dominant wavelength is around $\lambda=L/6 \approx 40$ km, while at $t\sim47$ sec it becomes $\lambda=L/2$ and finally after $t\sim100$ sec the dominant wavelength is equal to the whole size of the box in $X$ direction, L. Similar to the $\theta=90$\degree\ case, such behavior is a consequence of the non-linear interaction of flows, powering energy from small to large scales \citep[see, e.g.][]{Jun+etal1995b, Wang+Robertson1985}.

%%%%%%%%%%%%%%%%%%%%%%%%%%%%%%%%%%%%%%%%%%%%%%%%%%%%%%%%%%%%%%%%
\begin{figure}[!h]
\center
\includegraphics[height=12cm]{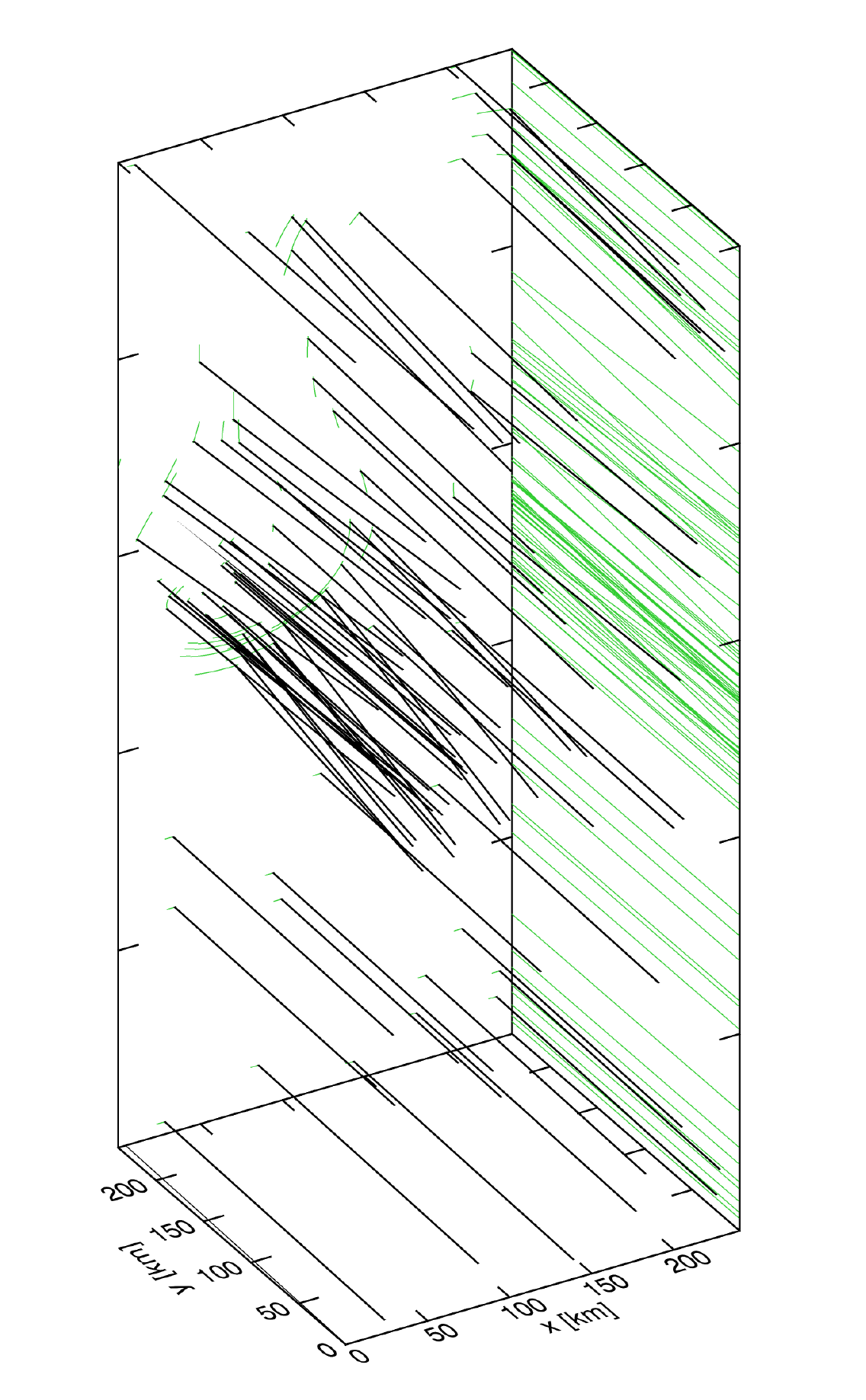}
\caption{{\footnotesize Three-dimensional representation of the magnetic field lines in a snapshot of the AD simulation with $\theta=89$\degree\ at $t=160$ sec. The vertical $Z$ axis is 1000 km long. The density of the field lines is proportional to the magnetic field strength and their origins in $XZ$ plane is random. Projections of the field lines into $XZ$ and $YZ$ planes are indicated in green color. } }\label{fig:b3d}
\end{figure}
%%%%%%%%%%%%%%%%%%%%%%%%%%%%%%%%%%%%%%%%%%%%%%%%%%%%%%%%%%%%%%%%

Several interesting phenomena happen during the evolution of the RTI in Fig.~\ref{fig:tevol89}. After being formed at about 60-70 sec of the simulation, the big drop starts falling at a speed of about 3--5 \kms. Similar range of velocities are also found in the simulations of \citet{Hillier+etal2012a, Hillier+etal2012b}. On its way down the drop deforms the magnetic field lines, compressing them. One can get a false impression from the figure that the magnetic field lines are significantly twisted and even become vertical. This is not so since the magnetic field lines are actually outside of the plane $XZ$, directed towards us. Fig.~\ref{fig:b3d} provides the three-dimensional rendering of the magnetic field lines in a single snapshot, showing that, as the drop evolves, the magnetic field lines are piled up together and slightly inclined and twisted, otherwise remaining essentially horizontal.

\begin{figure*}
\center
\includegraphics[width=14cm]{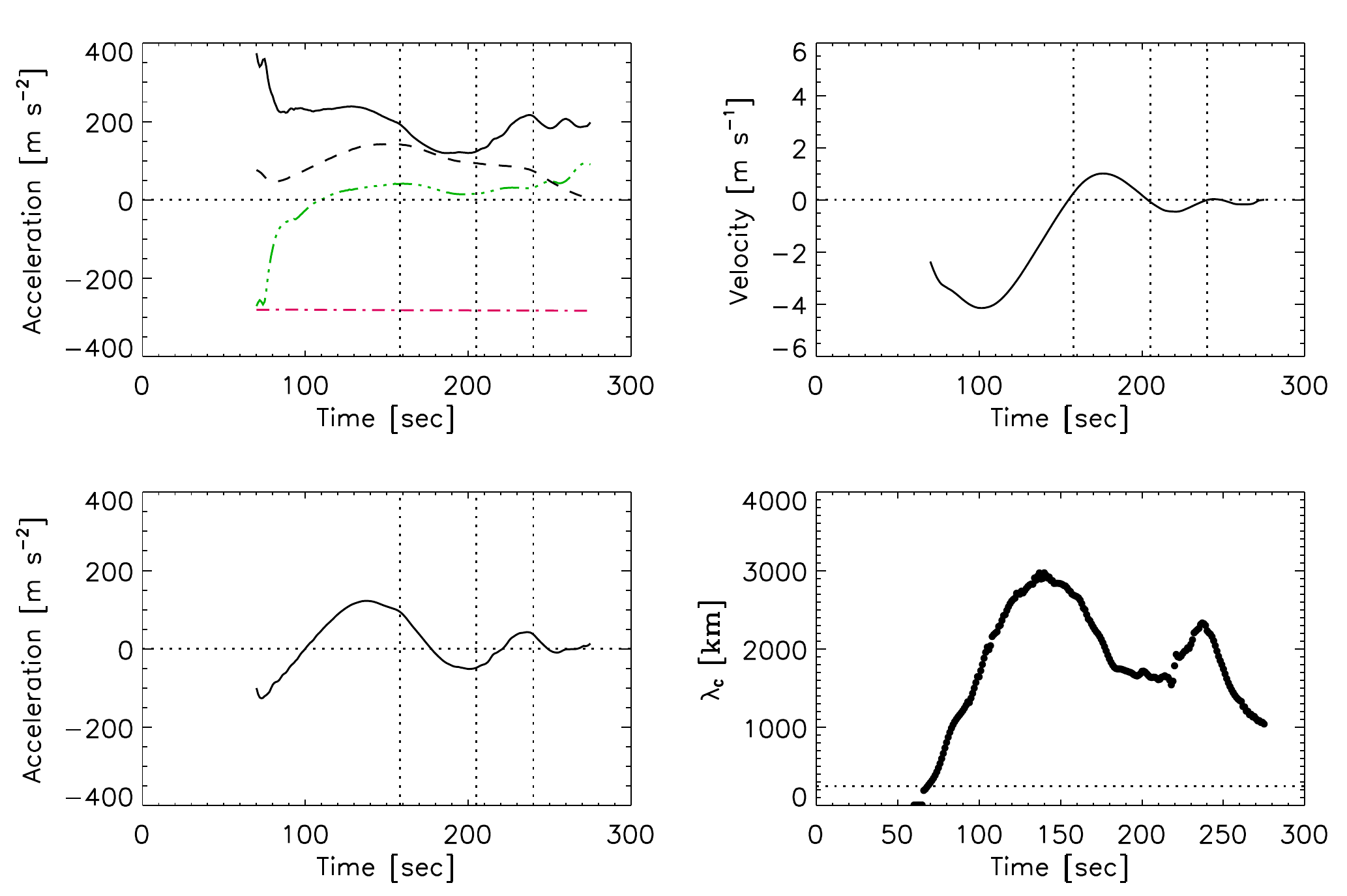}
\caption{{\footnotesize {\it Top left}: acceleration as a function of time due to the gravity force (dashed-dotted red line), gas pressure gradient force (dashed triple dotted green line), magnetic pressure gradient force (solid black line) and magnetic tension force (dashed black line). The values are integrated over the big drop from Fig.~\ref{fig:tevol89} (AD run) after it has been formed at about 70 sec of the simulation time. {\it Bottom left}: acceleration due to the sum of all above forces. {\it Top right}: vertical velocity of the big drop. Vertical dotted lines indicate time instants of zero velocity. {\it Bottom right}: effective cut-off wavelength calculated from Eq.~\ref{eq:lambdac} for the horizontal field component $B\cos(\theta)$ at the bottom of the drop. Horizontal dotted line indicates the size of the computational domain $L=250$ km.} }\label{fig:vdrop}
\end{figure*}

\begin{figure*}
\center
\includegraphics[width=14cm]{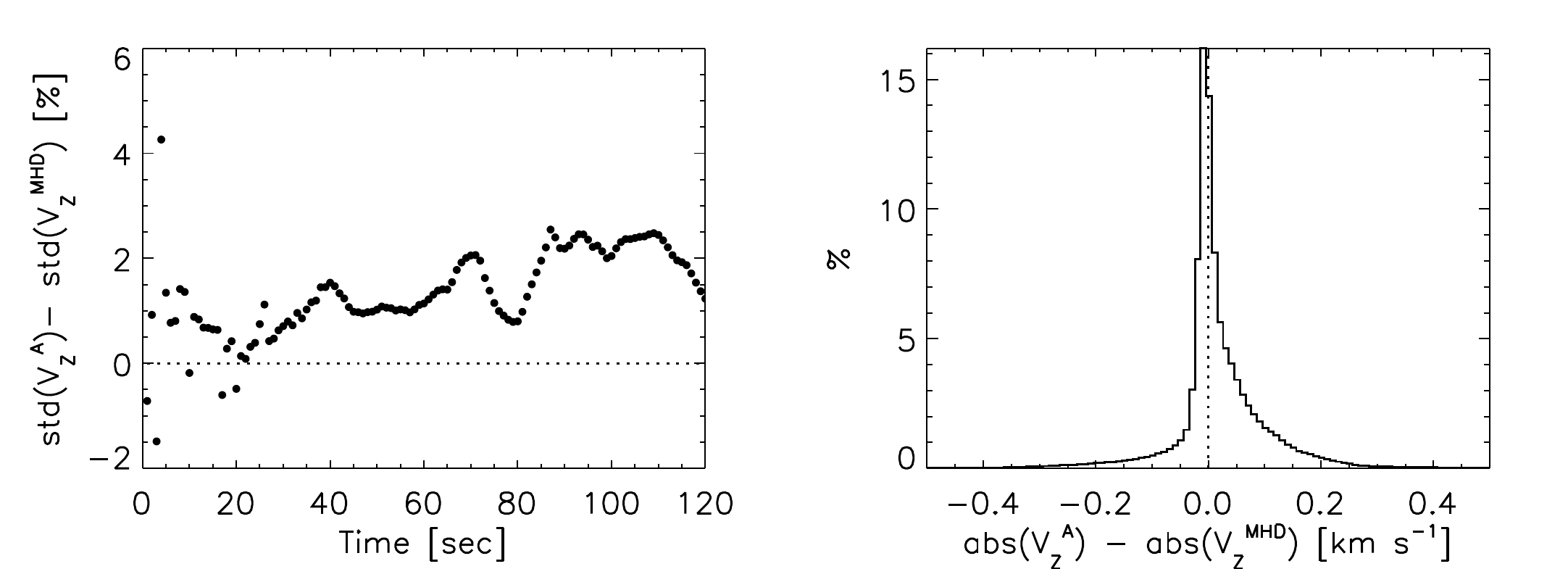}
\caption{{\footnotesize {\it Left}: Difference between the standard deviations of the vertical velocity near the discontinuity in the  AD and MHD $\theta=89$\degree\ simulations during the first 120 sec. {\it Right}: histogram of the difference between the absolute values of the vertical velocity near the discontinuity in the AD and MHD $\theta=89$\degree\ simulations during the first 120 sec. }
}\label{fig:vzlinear89}
\end{figure*}

\begin{figure*}
\center
\includegraphics[width=14cm]{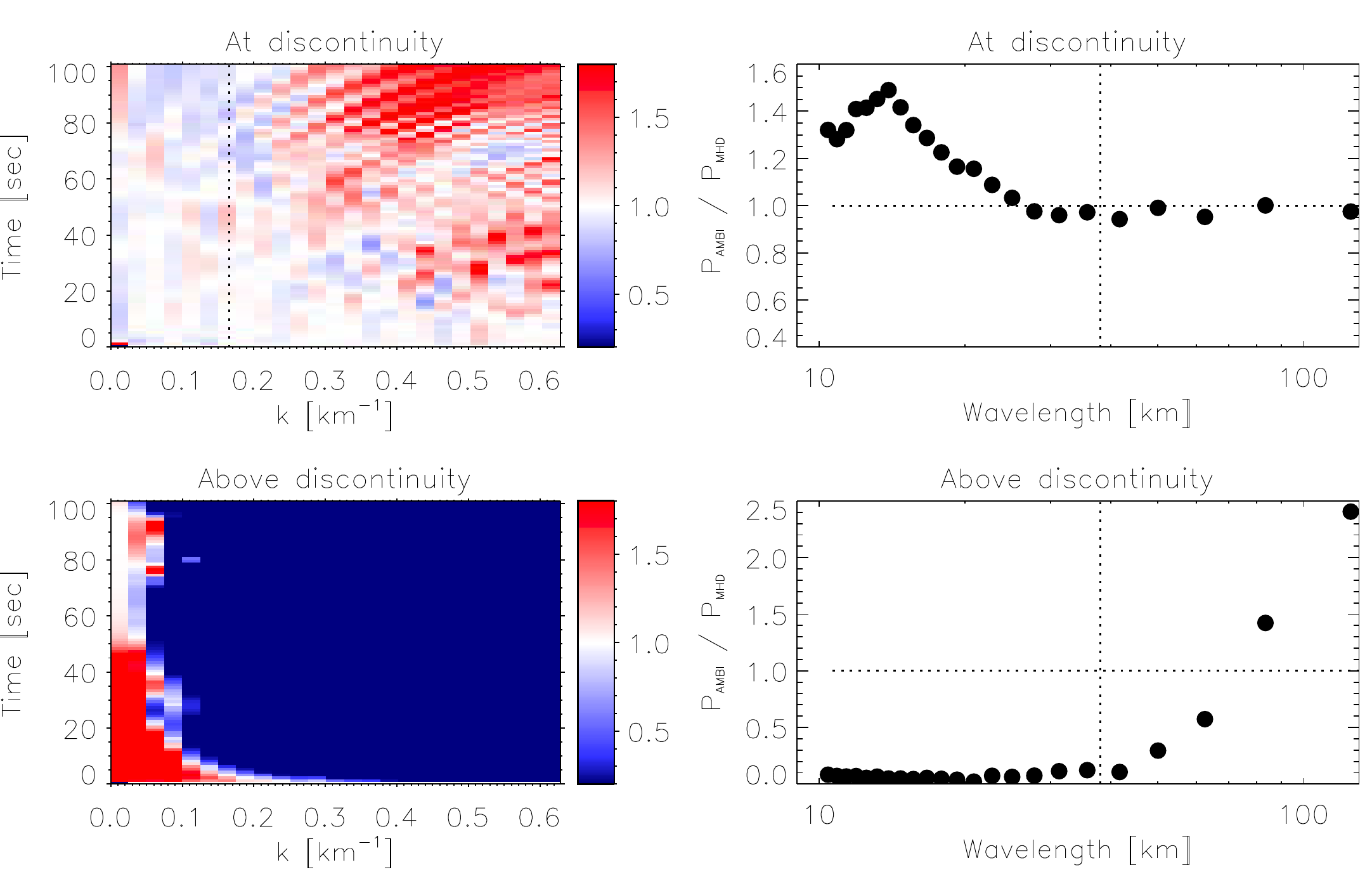}
\caption{{\footnotesize Fourier analysis of scales developed in the $\theta=89$\degree\ simulations. The format of the figure is the same as Fig.~\ref{fig:scales90}. The vertical dotted lines at the right panels marks the cut-off wavelength $\lambda_c=38$ km.} }\label{fig:scales89}
\end{figure*}

\begin{figure*}
\center
\includegraphics[width=18cm]{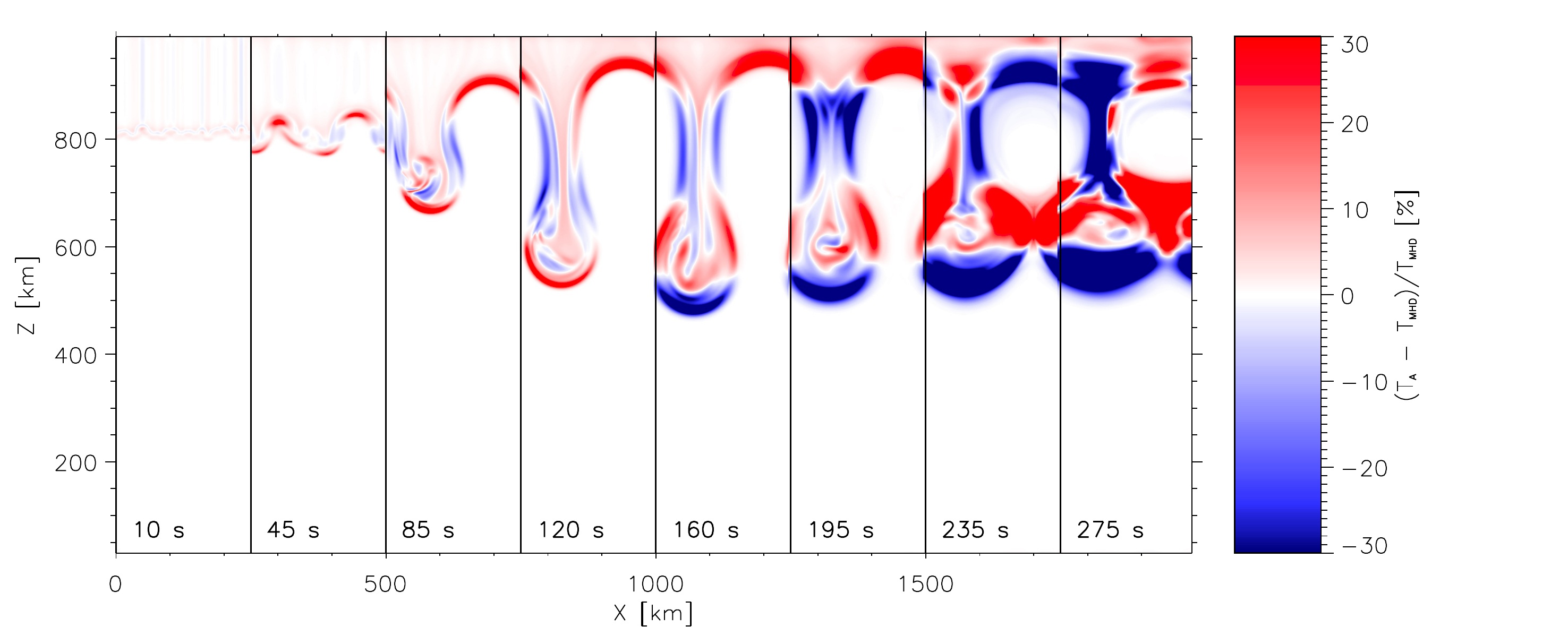}
\caption{{\footnotesize Difference between the temperature variations in the AD and MHD simulations with $\theta=89$\degree, $(T_{\rm A}-T_{\rm MHD})/T_{\rm MHD}$. The size of each snapshot is 250$\times$1000 km, the elapsed time is given at the bottom of each panel. Red colors mean that the AD model is hotter.} }\label{fig:dt89}
\end{figure*}

The time evolution of the velocity averaged over the drop is shown at the upper right panel of Fig.~\ref{fig:vdrop}. We observe that after $t\sim100$ sec the downward motion of the drop is slowed down and stops at $t\sim160$ sec. Then the drop starts rising again, extending in the horizontal direction (Fig.~\ref{fig:tevol89}). It remains oscillating around the equilibrium position and the amplitude of this oscillation decreases in time, until the motion finally ceases. The forces responsible for this behavior are given at the left panels of Fig.~\ref{fig:vdrop}. At the beginning of the evolution, the magnetic pressure gradient force almost balances the gas pressure gradient one, the magnetic tension force is small, and the gravity force dominates and causes the downward acceleration of the drop. The magnetic pressure gradient force remains the leading one to compensate the gravity over the whole evolution. The increase of the magnetic pressure is produced as the drop compresses the magnetic field lines on its fall. The variations of the magnetic pressure essentially cause the oscillation of the total acceleration (bottom left panel) and of the velocity (upper right panel). At the end of the evolution both gas pressure and magnetic pressure gradient forces act together to balance the gravity of the drop and stop its motion. Note that the force equilibrium produced in this simulation is different to the one found in \citet{Hillier+etal2012b} where the main force to compensate the gravity was the magnetic tension force. Such difference is caused by the different initial equilibrium of the system. In our case, the magnetic field is initially homogeneous and does not exert any force on the plasma, while in the Kippenhahn-Schl{\"u}ter  model of the prominence adopted by\citet{Hillier+etal2012b} the plasma is supported against gravity by the magnetic tension.

Besides the force balance, an increase of the horizontal field component below the drop causes an increase of the effective cut-off wavelength of the instability. The bottom right panel of Fig~\ref{fig:vdrop} shows the $\lambda_c$ estimated from Eq.~\ref{eq:lambdac} taking the values of the horizontal field component, $B\cos(\theta)$ immediately below the drop. One can see that $\lambda_c$ quickly increases and becomes larger than the size of the domain, $L$, already at $t\sim80$ sec. Though $\lambda_c$ given by Eq.~\ref{eq:lambdac}  is only a qualitative estimate, it indicates that perturbations with wavelength below $L$ can not grow further.

%%%%%%%%%%%%%%%%%%%%%%%%%%%%%%%%%%%%%%%%%%%%%%%%%%%%%%%%%%%%%%%%
%%%%%%%%%%%%%%%%%%%%%%%%%%%%%%%%%%%%%%%%%%%%%%%%%%%%%%%%%%%%%%%%

%%%%%%%%%%%%%%%%%%%%%%%%%%%%%%%%%%%%%%%%%%%%%%%%%%%%%%%%%%%%%%%%
%%%%%%%%%%%%%%%%%%%%%%%%%%%%%%%%%%%%%%%%%%%%%%%%%%%%%%%%%%%%%%%%

%%%%%%%%%%%%%%%%%%%%%%%%%%%%%%%%%%%%%%%%%%%%%%%%%%%%%%%%%%%%%%%%
%%%%%%%%%%%%%%%%%%%%%%%%%%%%%%%%%%%%%%%%%%%%%%%%%%%%%%%%%%%%%%%%

The histograms of the velocities in the simulation of $\theta=89$\degree\ are similar to that shown in Fig.~\ref{fig:vz90} except that the range of the velocity variations is smaller, $\pm10$ \kms. We do not show the corresponding figure. Instead, we investigate if there are systematic differences between the velocity field in the AD and MHD simulations. The structures that develop in both simulations are rather similar (unlike the turbulent flows in the $\theta=90$\degree\ simulation), therefore we can directly compare the snapshots of the vertical velocity in both cases. The left panel of Fig.~\ref{fig:vzlinear89} gives the difference between the standard deviations of the vertical velocity measured around the discontinuity in the AD and the MHD runs. The difference is measured in percents from the value in the MHD case. From the very beginning of the instability the standard deviation of the velocity in the AD run results to be 2--3\% larger. The difference is not large but it is systematic, and the velocities in the AD case are slightly larger. A similar conclusion is obtained from the right panel of Fig.~\ref{fig:vzlinear89} where we show the histogram of the difference in the absolute velocity values around the discontinuity between the AD and the MHD run. The histogram shows the prevalence of positive values (62\%) so that the absolute value of the velocity is larger in the AD simulation, i.e. the instability develops slightly faster in this case.

To evaluate the responsible scales and to demonstrate the larger growth rate in the AD run we perform the same Fourier analysis as in Fig.~\ref{fig:scales90}, but for the $\theta=89$\degree\ simulation. The results are given in Fig.~\ref{fig:scales89}. Similar to Fig.~\ref{fig:scales90} we divide the simulation domain into two zones, at the discontinuity and above it. The power ratio in the prominence part of the domain above the discontinuity behaves similar to the $\theta=90$\degree\ run, the ambipolar diffusion term acting as pure diffusion and quickly removing small scales. At the discontinuity, the power ratio is significantly different from the $\theta=90$\degree\ case. Now there is a clear increase in the growth rate of the small-scale harmonics in the AD simulation compared to the MHD one. The growth rate of the large-scale harmonics is the same in both cases. The change in the behavior happens at about $\lambda \sim 30$ km, as indeed expected, because this value is close to the cut-off wavelength $\lambda_c$. This result confirms and extends the conclusion from the linear theory \citep{Diaz+etal2013}, see Fig.~\ref{fig:linear}. The solid line in Fig.~\ref{fig:linear} gives the growth rate from the linear theory for the parameters of our model and field orientation of $\theta=89$\degree, taking the ambipolar diffusion term into account. Indeed, compared to the MHD case (dashed line in Fig.~\ref{fig:linear}), all RTI modes become unstable when the presence of neutral atoms is accounted for in the analysis. We obtain up to 50\% increase of the small-scale harmonics growth rate, in good agreement with the linear analysis.

Fig.~\ref{fig:dt89} shows the temperature difference between the snapshots of the AD and MHD simulations. It follows that appreciable temperature differences exist at all stages of evolution of the RTI. Initially, there are larger temperatures at the discontinuity and in the prominence part of the domain in the AD case due to current dissipation and heating \citep[see ][]{Khomenko+Collados2012}. The interior of the drop is maintained about 5--10\% hotter in the AD at times up to 160 sec. After the initial stage, the neutral and ionized material is mixed, the shape of the drop starts to be more and more different in both runs and this causes a large temperature difference at the transition between the drop and the corona. All in all, the temperature differences are not small, reaching at maximum $\sim$60 kK, or about 30\% of the actual value.

\subsection{Ion-neutral drift}
\label{sect:drift}

The relative velocity between the neutral and the ionized plasma components (drift velocity) can be used as a parameter for evaluating the importance of non-ideal plasma effects due to the presence of neutrals. An approximate expression for this velocity can be derived by combining the momentum equation for electrons, ions and neutrals and neglecting the electron inertia and the time derivative of the ion-neutral drift velocity:

\begin{equation}
\label{eq:w}
\vec{w} = \vec{u}_i - \vec{u}_n=\frac{\xi_n}{\alpha_n} \left[\vec{J}
\times\vec{B} \right] - \frac{\vec{G}}{\alpha_n} + \epsilon\frac{\vec{J}}{en_e}+\frac{\xi_n\rho_e}{\alpha_n}\vec{g}
\end{equation}
For the derivation of Eq.~\ref{eq:w}, see, e.g., \citet{Diaz+etal2013}. The meaning of the variables is the following:
\begin{equation}
\vec{G}=(2\xi_n \vec{\nabla}p_e - \xi_i \vec{\nabla} p_n)
\end{equation}
is the partial pressure gradient term; $\xi_n=\rho_n/\rho$ and $\xi_i=\rho_i/\rho$ are neutral and ion fraction, correspondingly; $p_e$ and $p_n$ are the electron and neutral pressure, respectively, and $\alpha_n$ is the neutral collisional frequency, calculated from the following expression:
\begin{equation}
\alpha_n=m_n n_i\nu_{in} + m_e n_i \nu_{en}
\end{equation}
where $n_i$ is ion number density and the collisional frequencies $\nu_{in}$ and $\nu_{en}$ are given by Eq.~\ref{eq:nus}. The small parameter $\epsilon$ is defined as:
\begin{equation}
\epsilon=\frac{\rho_e\nu_{en}}{\alpha_n} \approx 0.003
\end{equation}

Keeping in mind all approximations used to derive Eq.~\ref{eq:w}, once the drift velocity is obtained, we can recover the individual velocities of ions and neutrals from the linear system of Eqs.~\ref{eq:u} and \ref{eq:w}:
\begin{eqnarray} \label{eq:moms}
\vec{u}_n &=&\vec{u} + (\xi_n-1)\vec{w} \nonumber \\
\vec{u}_i &=&\vec{u} + \xi_n\vec{w}
\end{eqnarray}

The last two terms in Eq.~\ref{eq:w} are small, and we keep for the analysis only the two leading ones. The drift velocity defined this way is produced by cross-field currents (first term) and by gradients of the partial pressures (second term). A large drift velocity means that the species are not entirely coupled together by collisions.

%%%%%%%%%%%%%%%%%%%%%%%%%%%%%%%%%%%%%%%%%%%%%%%%%%%%%%%%%%%%%%%%
%%%%%%%%%%%%%%%%%%%%%%%%%%%%%%%%%%%%%%%%%%%%%%%%%%%%%%%%%%%%%%%%

\begin{figure*}
\center
\includegraphics[width=18cm]{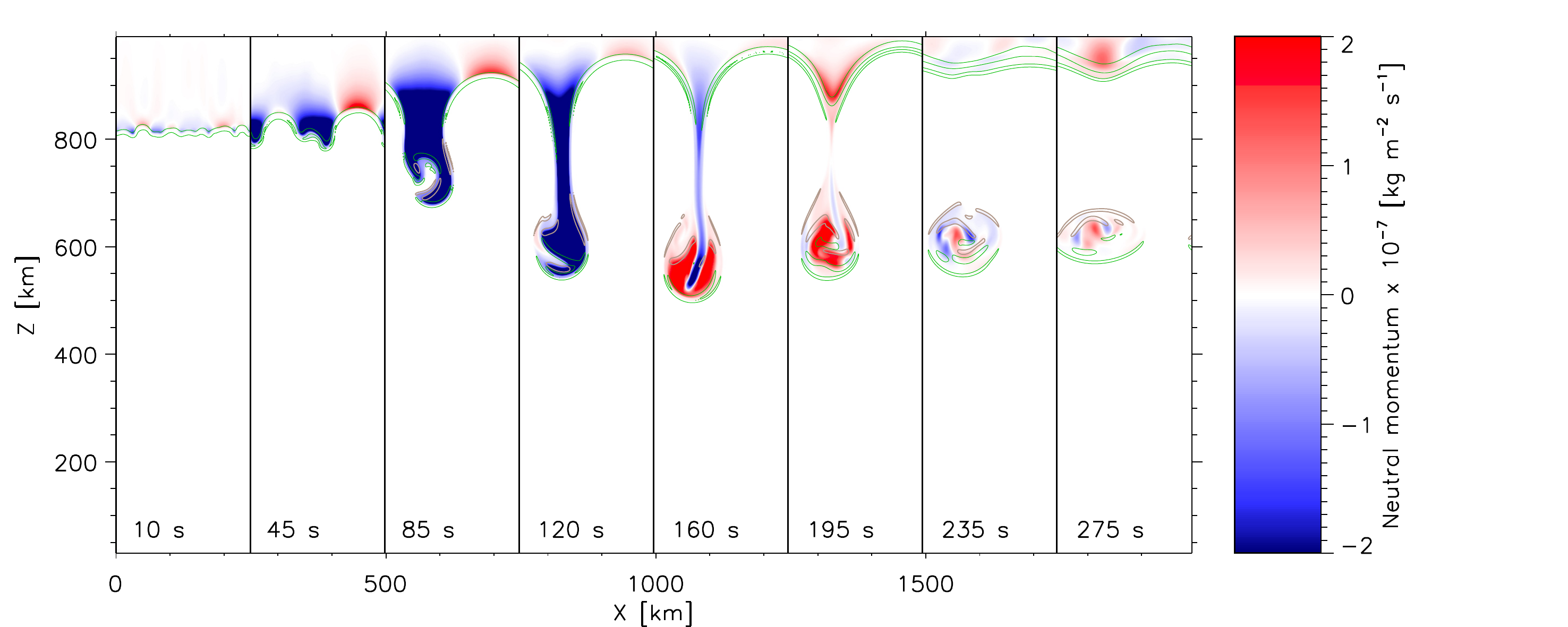}
\includegraphics[width=18cm]{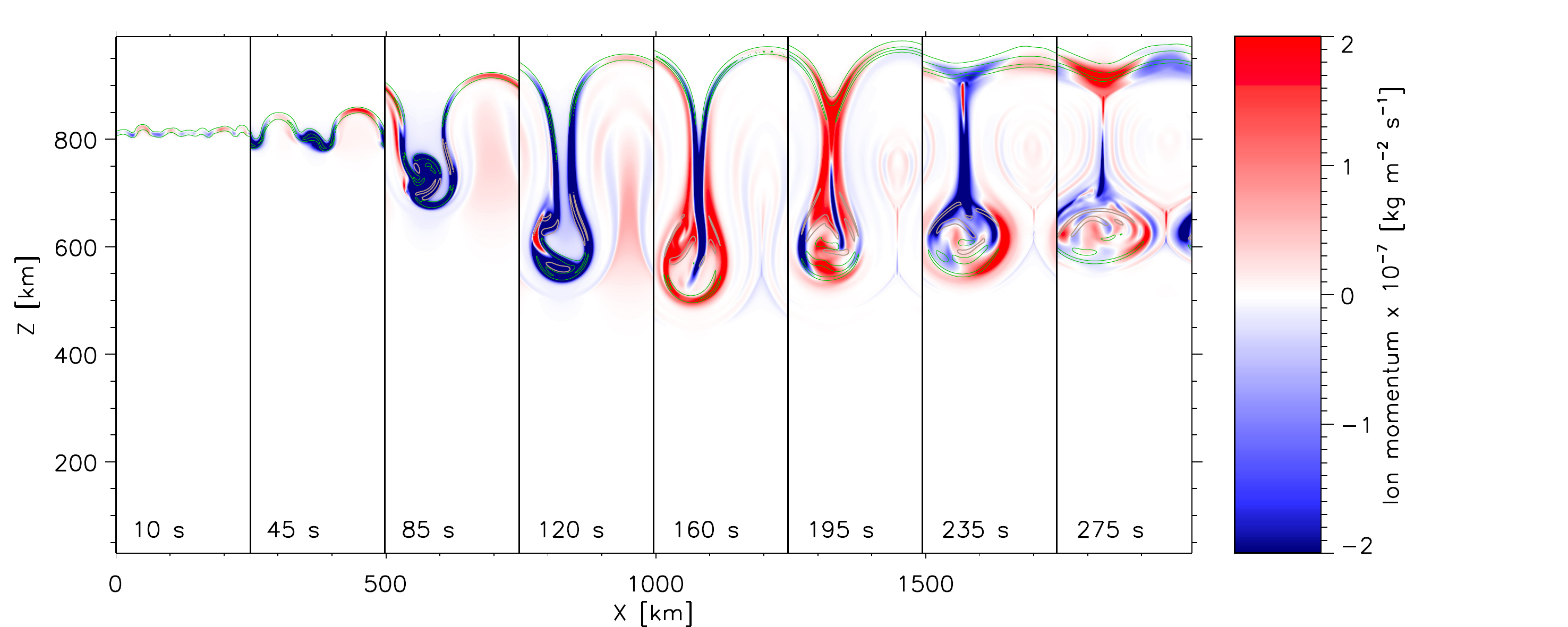}
\caption{{\footnotesize {\it Top}: time evolution of the vertical component of the neutral momentum from Eq.~\ref{eq:moms} in the $\theta=89$\degree\ AD run. Negative values mean downward motion. The contours of vertical drift momentum $\sqrt(\rho_i\rho_n) w_z=10^{-9}$ and $-10^{-9}$ kg m$^{-2}$s$^{-1}$, Eq.~\ref{eq:momw}, are overplotted in green and grey colors, respectively. {\it Bottom:} same for the ion momentum. The size of each snapshot is 250$\times$1000 km, the elapsed time is given at the bottom of each panel, the color scheme is constant in time and is the same in both panels. } }\label{fig:omega89}
\end{figure*}

The individual velocities of species do not carry much information themselves, since they do not reflect the amount of material moving with a given velocity. Such information is contained in their momenta:
\begin{equation}
\vec{p}_n=\rho_n\vec{u}_n; \,\,\, \vec{p}_i=\rho_i\vec{u}_i.
\end{equation}
For this reason, following \citet{Pandey+Wardle2008}, we define the ion-neutral drift momentum:
\begin{equation} \label{eq:momw}
\vec{p}_D=\sqrt(\rho_i\rho_n)\vec{w}.
\end{equation}
To preserve the sign of the drift momentum we do not use the square of this quantity, unlike, e.g., \citet{Pandey+Wardle2008} and \citet{MartinezSykora+etal2012}. We have calculated the neutral, ion, and drift momenta a posteriori from the parameters of the AD $\theta=89$\degree\ simulation and the result is displayed in Fig.~\ref{fig:omega89}.

At the beginning of the evolution ($t=10$ sec), the upper prominence part of the domain has a very low ionization fraction, and the lower coronal part is completely ionized, but the absolute value of density in the corona is about 100 times lower than in the prominence. Significant velocities appear only at the interface and, as a consequence, both ion and neutral momenta are only significant at the PCTR. Later on ($t=45-120$ sec) there appears significant negative neutral momentum in the prominence part of the domain, pushing the drop downwards. The sharp transition in the ionization fraction between the prominence and the corona disappears, producing a smooth increase of the ionization fraction at the interface and negative ion momentum just at the border of the drop. At times $t=160-195$ sec the negative neutral momentum ceases, and there is a positive ion momentum surrounding the drop and pushing it upwards. Later, at $t=235-275$ sec, the ionized and the neutral fluids inside the drop mix up, and the neutral momentum becomes small due to the increase of the ionization fraction inside the drop.

%%%%%%%%%%%%%%%%%%%%%%%%%%%%%%%%%%%%%%%%%%%%%%%%%%%%%%%%%%%%%%%%
%%%%%%%%%%%%%%%%%%%%%%%%%%%%%%%%%%%%%%%%%%%%%%%%%%%%%%%%%%%%%%%%

The contours in Fig.~\ref{fig:omega89} show the time evolution of the drift momentum calculated according to Eq.~\ref{eq:momw}. This quantity is large at the PCTR. The absolute values of the drift momentum reach 1--2$\times 10^{-8}$ kg m$^{-2}$s$^{-1}$ which makes about 10--15\% of the values of the individual ion and neutral momenta. At the first instants of the instability $\vec{p}_D$ is positive. Combined with the negative ion and neutral momenta, positive values of $\vec{p}_D$ mean that neutrals fall faster than the ions. The drift momentum remains positive at the bottom part of the drop at all instants. At $t=160$ s, when the drop starts rising, both ion and neutral momenta are positive, and $\vec{p}_D>0$ below the drop mean faster upward motion of ions. Such behavior can be expected since the magnetic field does not act on neutrals directly, but only via the collisions with charged particles. In the limiting case of the absence of collisions, the instability in the neutral gas would develop entirely hydrodynamically, i.e. without cut-off wavelength. The collisional interaction with ions prevents this from happening and forces the collective behavior of the plasma. Nevertheless, the coupling becomes less strong at the PCTR, where strong gradients in all parameters, including the collisional frequency, are present. The non-ideal partial ionization effects on the RTI essentially originate from this narrow layer.

\section{Discussion and Conclusions}

We have studied how the presence of neutral atoms in a prominence plasma influences the development of the Rayleigh--Taylor instability at the coronal-prominence interface by means of numerical 2.5D simulations. Our approach consisted in solving the single-fluid quasi-MHD equations including physical diffusion term (ambipolar diffusion) in the induction and energy equations. Such approach is justified in the regime of strong collisional coupling of the plasma. Our main goal was to investigate the RTI in partially ionized plasmas in the non-linear regime to verify and extend the conclusions from the linear theory.

Our main finding is that the configuration containing neutral atoms is always unstable on small scales. While in the completely ionized plasma the growth rate of the small-scale harmonics is lowered (or becomes zero) due to the action of magnetic forces, this is not like that if neutral atoms are present. We obtain an increase in the growth rate of the small scale harmonics, up to 50\%, when partial ionization effects are taken into account (Fig.~\ref{fig:scales89}). This result is in good agreement with the linear theory \citep[][see our Fig.~\ref{fig:linear}]{Soler+etal2012, Diaz+etal2012,  Diaz+etal2013}. We show that, relaxing the approximations of the linear analysis, the growth rate is still large in the non-linear regime of the RTI. A statistical comparison reveals that this fast growth rate at small scales produces, on average, a 2--3 percent larger flow velocities in the model with  ambipolar diffusion term ``on'' compared to the purely MHD model (Fig.~\ref{fig:vzlinear89}).

Another action of the ambipolar diffusion is the dissipation of currents in the direction perpendicular to the magnetic field. Such dissipation allows to transform magnetic energy into internal energy \citep{Khomenko+Collados2012, MartinezSykora+etal2012}, and this results in heating of the plasma in the prominence part of the domain. The temperature increase due to this effect is about 5--10\% inside the prominence, and about 10--20\% at the PCTR, compared to the model without ambipolar diffusion (Fig.~\ref{fig:dt89}). While the heating is produced, we observe that the amplitude of perpendicular currents, $J_{\perp}$, becomes progressively lower in the model with ambipolar diffusion, reaching 20--30\% difference with the pure MHD case. The ambipolar diffusion introduces an anisotropy in the system since, with time, perpendicular currents get dissipated and the longitudinal ones are unaltered. Such anisotropic dissipation tends to create force-free structures, as was shown in the simulations by \citet{Leake+Arber2006, Arber2009}.

As the instability evolves, the initially sharp interface gets smoother and is filled with a mixture of coronal and prominence material. In this transition layer, the density, ionization fraction, collisional frequency, and other parameters vary strongly from prominence to coronal values. Due to the decrease of density, the collisional coupling becomes less strong. As a consequence of that, a non-negligible drift momentum appears at this layer (Eq~\ref{eq:momw}, Fig.~\ref{fig:omega89}), with a value of 10--15\% of the individual ion and neutral momenta. The sign of the drift momentum indicates that the neutral atoms at the bottom part of the downflowing drop move with slightly larger downward velocities. The neutrals feel the presence of the magnetic Lorentz force only through the collisions with the ionized particles and, once the collisional coupling weakens, this relative motion between the components becomes possible. Perpendicular currents also reach maximum values inside the transition layer between the prominence and coronal material. Therefore, the action of partial ionization effects on the RTI is localized in space to a small zone. This explains why the inclusion of these effects in our current modeling only alters the global parameters of the flow (thermodynamic parameters, velocities, magnetic field) by no more than a few tens of percent. In a different, non current-free equilibrium configuration, the action of partial ionization effects could be significantly amplified.

Our initial equilibrium configuration is the simplest possible, purely hydrodynamical with a homogeneous magnetic field. Such configuration is different from the equilibrium usually thought to exist in prominences producing their large-scale stability \citep{Tandberg-Hanssen1995, Mackay+etal2010}. The support of the prominence material is thought to be provided by the magnetic tension. Contrarily, we neglect the effects of the magnetic field curvature in our analysis. Our choice was motivated by two reasons. On the one hand, such equilibrium is a natural choice for the 2.5D modeling that does not allow to include the effects of magnetic field curvature perpendicular to the perturbation plane \citep[as Kippenhahn-Schl{\"u}ter model, see][]{Hillier+etal2012a, Hillier+etal2012b}. We have presented here an exploratory study to investigate the importance of partial ionization effects, and 2.5D simulations, instead of full 3D, allow for faster calculations. On the other hand, we have also pursued with this work an adequate comparison with the linear theory, where the same equilibrium is adopted \citep{Diaz+etal2013}.

Our equilibrium model causes several limitations. Since the temperature in the prominence part of the domain is rather low, the pressure scale height is only a few hundreds of km, and we are unable to extend the model in height to have a large slab of prominence material, since the pressure and density quickly drop to coronal values. This limits the size of the computational box, making impossible the direct comparison to observations of the whole prominence structure. The current-free equilibrium is another drawback, since, as already mentioned above, the action of the partial ionization effects is only limited to a narrow transition zone, decreasing its potential influence.

As a consequence of our equilibrium model, the force balance developed in the simulations is different from the one found in \citet{Hillier+etal2012a, Hillier+etal2012b}. The main force to balance gravity in the downflowing drop in Fig.~\ref{fig:tevol89} is found to be the magnetic pressure force (Fig.~\ref{fig:vdrop}), unlike the magnetic tension force in the aforementioned papers. Since we have not introduced any buoyant rising material, as in \citet{Hillier+etal2012a, Hillier+etal2012b}, the distribution of upflows and downflows is significantly more symmetric in our case (Fig.~\ref{fig:vz90}). However, due to mass conservation, the upflowing rising bubbles of the coronal material have significantly larger sizes than the downflowing fingers of dense prominence material (Fig.s~\ref{fig:tevol90} and \ref{fig:tevol89}). Therefore, assuming the density to be a proxy for the intensity in $H_{\alpha}$ imaging observations, observational detection of upflows is easier and could cause the observed asymmetry \citep{Berger2010, Ryutova+etal2010}.

The simulations described above are only done for two orientations of the magnetic field, one normal to the perturbation plane, $\theta=90$\degree, and another one skewed by just one degree from the normal, $\theta=89$\degree. It might seem surprising that the structures developed in both cases are so different. Such behavior is nevertheless expected from the properties of the non-linear flow cascade and are observed in many other simulations of the RTI in different astrophysical contexts in 2 and 3 dimensions \citep{Jun+etal1995b, Wang+Robertson1985, Isobe+etal2006, Stone+Gardiner2007a, Stone+Gardiner2007b, Hillier+etal2012a}. The increase of the dominant scale with time (Fig.~\ref{fig:maxscale}) is in a good agreement with other works \citep{Jun+etal1995b}.  Since the maximum perturbation wavelength was of the size of the computational domain, $\lambda=$250 km, it is natural that with time the perturbation of this scale dominates. The results of our initial calculations in a larger box \citep{Khomenko+etal2013} show that the same perturbation develops several drops on the same time scale. There, a simulation with $\theta=88$\degree\ was also analyzed and essentially lead to the same conclusions as for the enhanced growth rate of small scale modes with the ambipolar diffusion term ``on''. The disappearance of small scales in the $\theta=89$\degree\ simulation is caused by the cut-off wavelength introduced by the component of the magnetic field in the perturbation plane and is also in agreement with other numerical works \citep{Stone+Gardiner2007a}. One also has to keep in mind that the size of our computational box is relatively small compared to the cut-off wavelength introduced by magnetic field skewed by just 1\degree\ away from normal, $\lambda_c=38$ km.

The main issue of our modeling is the use of the Saha equation to update the electron density and neutral fraction at each time step of the simulations. At prominence temperatures of the plasma, the deviations from the instantaneous ionization equilibrium can be significant, and the use of Saha equation may lead to an underestimation of the electron density. A more appropriate approach would be to consider a time-dependent ionization balance. \citet{Leenaarts2006, Leenaarts2007, Wedemeyer2011} have shown that, while the ionization process happens rapidly, the recombination is slow, so the ionization fraction is maintained rather constant in time and space even when significant temperature fluctuations are present. The calculations of the impact of time-dependent ionization balance on the development of the RTI in partially ionized plasma needs a thorough study. However, such calculations require significant computational resources and are beyond the scope of the present explorative work.

Besides the Rayleigh-Taylor instability per se, the posterior time evolution of the downflowing drop from Fig.~\ref{fig:tevol89} deserves a separate discussion. After the downward motion of the drop stops at $t\approx160$ sec, it remains oscillating around the equilibrium position, extending in the horizontal direction. The evolution of the drop brings close the magnetic field lines at the coronal part of the domain and reconnection happens at about 200 sec of the simulation.  The plasmoid formed after this reconnection is visible at time 237 sec in Fig.~\ref{fig:tevol89}. The layer of plasma linking the drop to the main part of the prominence becomes thinner and finally another reconnection happens in the chromospheric part of the domain. Another magnetic island is formed at the location of the drop, and the drop becomes almost completely disconnected from the rest of the prominence, extending even more in the horizontal direction and forming a thread (not shown in Fig.~\ref{fig:tevol89}). The reconnections and the formation of the horizontal thread will need further analysis in a separate paper. The process of the drop falling across the horizontal magnetic field lines under the action of gravity, its disconnection from the magnetic field of the rest of the prominence and forming a finite plasma island  may resemble the mechanism proposed by \cite{Haerendel2011}.

Summarizing all above, we conclude that partial ionization effects on the Rayleigh-Taylor instability in prominences are non-negligible and have to be taken into account in models of prominence dynamics. In the future, we will consider larger simulation boxes in three dimensions to perform the comparison with observations, and will investigate the development of the instability for different ionization fractions and initial equilibrium configurations.

\begin{acknowledgements}
This work is partially supported by the Spanish Ministry of Science through projects AYA2010-18029 and AYA2011-24808. This work contributes to the deliverables identified in FP7 European Research Council grant agreement 277829, ``Magnetic connectivity through the Solar Partially Ionized Atmosphere''.
\end{acknowledgements}

%%%\aareferences

\appendix
\section{Notes on the nomenclature}

%The collisions of charged particles with neutrals allow the magnetic field and plasma to slip with respect to each other, just as c%collisions among the charged particles alone do so.  Both plasma/plasma and plasma/neutral collisions act to break the frozen-%in flux condition and allow the magnetic field to diffuse through the plasma, as described by the Spitzer and Cowling resistivity %terms, respectively.

The term proportional to $\eta_A\vec{J}_\perp$ in Eqs. \ref{eq:system} describes the coupling of neutral particles  to the charges. It causes the magnetic field to diffuse through neutral gas due to collisions between neutrals and charged particles. In an astrophysical context, especially in the fields of interstellar medium, star formation and stellar atmosphere modeling, this  process is usually referred to as  ``ambipolar diffusion''. This notation was first used by Spitzer in his books ``Diffuse matter in space'' (\citeyear{Spitzer1968}) and ``Physical processes in the interstellar medium''  (\citeyear{Spitzer1978}). Ambipolar diffusion makes possible the star formation from gravitational collapse in magnetized protostellar clouds, as was suggested in the pioneering work by \citet{Mestel+Spitzer1956}. In the subsequent years this terminology was adopted as a standard one in this field. An incomplete list of works where the decoupling of neutral particles from plasma is called ``ambipolar diffusion'' includes  \citet{Osterbrock1961, Nakano1977, Mouschovias1977, Mouschovias1981, Black+Scott1982, Scott1983, Paleologou1983, Zweibel1988, Myers1988, Fontenla1990, Dudorov1991, Babel1991, Fiedler1992, Ciolek1993, Brandenburg+Zweibel1994, Brandenburg+Zweibel1995, MacLow1995, Roberge1995, Smith1998, Greaves1999, Flower2003, Heitsch2004, Oishi2006, Li2006, Duffin2008, Stone+Zweibel2010, Bai+Stone2011, Jones+Downes2011}.

On the contrary, in the field of plasma physics, the term ``ambipolar diffusion'' refers to the diffusion of positive and negative particles at the same rate due to their Coulomb interaction, which maintains the charge neutrality at scales larger than the Debye length \citep[see standard plasma physics books by][]{Boyd+Sanderson, Krall+Trivelpiece}. The use of the same terminology applied to these two different phenomena may lead to confusion. 

In solar physics context, the additional diffusion due to neutrals is frequently referred to as ``Cowling resistivity'', following the notation by \citet{Braginskii1965} \citep[see the original work by][]{Cowling1956}. Strictly speaking, the Cowling resistivity includes both Ohmic and neutral diffusive terms, i.e. $\eta_C=\eta_A + \eta$, however, the Ohmic term is orders of magnitude lower in the solar atmosphere and is usually neglected. Recent citations where the ``Cowling resistivity'' terminology is used include, e.g., \citet{Khodachenko2004, Khodachenko2006, Leake+Arber2006, Forteza2007, Arber2007, Arber2009, Sakai+Smith2009}.  Alternatively, this term is also called ``Pedersen resistivity'' in works by \citet{Sakai2006, Krasnoselskikh2010, Leake+Linton2013, Leake+etal2013,  Cheung+Cameron2012}, and others. Nevertheless, many other works in solar physics use the terminology introduced by Spitzer, i.e. ``ambipolar diffusion'', see \citep{Chitre+Krishan2001, Pandey2008, Soler2009, Soler2010,  Hiraki2010, Singh2011, Cheung+Cameron2012, MartinezSykora+etal2012,  Soler+etal2012, Tsap2012, Diaz+etal2012, Khomenko+Collados2012}.

%Since the term ``ambipolar diffusion'' is widely used in the astrophysical literature, we consider it appropriate to follow the %common nomenclature through the current paper. 

In this paper we use ``ambipolar diffusion''  in the astrophysical sense, as a widely spread nomenclature in this research field.

\end{document}